\documentclass{article}
\usepackage{graphicx}
\usepackage{epstopdf}
\usepackage{amsmath}
\usepackage{geometry}
\usepackage{amsfonts}
\usepackage{amssymb}
\usepackage{color}
\usepackage{lscape}
\newcommand{\doublespacing}{\let\CS=\@currsize\renewcommand{\baselinesstrech}
	{2.0}\tiny\CS}
\begin{document}
	\newcommand{\bd}{\begin{document}}
		\newcommand{\ed}{\end{document}}
	\newcommand{\bc}{\begin{center}}
		\newcommand{\ec}{\end{center}}
	\newcommand{\bfr}{\begin{flushright}}
		\newcommand{\efr}{\end{flushright}}
	\newcommand{\lt}{\left}
	\newcommand{\rt}{\right}
	\newcommand{\vs}{\vspace}
	\newcommand{\hs}{\hspace}
	\newcommand{\beq}{\begin{equation}}
		\newcommand{\eeq}{\end{equation}}
	\newcommand{\lb}{\linebreak}
	\newcommand{\pb}{\pagebreak}
	\newcommand{\mb}{\makebox}
	\newcommand{\fb}{\framebox}
	\newcommand{\mc}{\multicolumn}
	\newcommand{\ben}{\begin{enumerate}}
		\newcommand{\een}{\end{enumerate}}
	\newcommand{\bit}{\begin{itemize}}
		\newcommand{\eit}{\end{itemize}}
	\newcommand{\oln}{\overline}
	\newcommand{\un}{\underline}
	\newcommand{\lefq}{\lefteqn}
	\newcommand{\ba}{\begin{array}}
		\newcommand{\ea}{\end{array}}
	\newcommand{\beqa}{\begin{eqnarray}}
		\newcommand{\eeqa}{\end{eqnarray}}
	\newcommand{\beqas}{\begin{eqnarray*}}
		\newcommand{\eeqas}{\end{eqnarray*}}
	\newcommand{\bfg}{\begin{figure}}
		\newcommand{\efg}{\end{figure}}
	\newcommand{\bds}{\begin{displaymath}}
		\newcommand{\eds}{\end{displaymath}}
	\newcommand{\btb}{\begin{tabbing}}
		\newcommand{\etb}{\end{tabbing}}
	\newcommand{\para}{\parallel}
	\newcommand{\pad}{\partial}
	\newcommand{\nn}{\nonumber}
	\newcommand{\la}{\leftarrow}
	\newcommand{\ra}{\rightarrow}
	\newcommand{\lgla}{\longleftarrow}
	\newcommand{\lgra}{\longrightarrow}
	\newcommand{\La}{\Leftarrow}\newcommand{\Ra}{\Rightarrow}
	\newcommand{\Lra}{\Leftrightarrow}
	\newcommand{\Lgla}{\Longleftarrow}
	\newcommand{\Lgra}{\Longrightarrow}
	\newcommand{\lan}{\langle}
	\newcommand{\ran}{\rangle}
	\renewcommand{\a}{\alpha}
	\renewcommand{\b}{\beta}
	\newcommand{\g}{\gamma}
	\newcommand{\G}{\Gamma}
	\renewcommand{\d}{\delta}
	\newcommand{\eps}{\epsilon}
	\newcommand{\Th}{\Theta}
	\newcommand{\s}{\sigma}
	\newcommand{\lam}{\lambda}
	\newcommand{\D}{\Delta}
	\newcommand{\ds}{\displaystyle}
	\newcommand{\vare}{E}
	\newcommand{\pr}{\prime}
	\newcommand{\ro}{\rho}
	\newcommand{\nab}{\nabla}
	\newcommand{\m}{\mu}
	\newcommand{\n}{\nu}
	\newcommand{\Sg}{\Sigma}
	\newcommand{\p}{\pi}
	\newcommand{\R}{I\!\!R}
	\newcommand{\om}{\omega}
	\newcommand{\Om}{\Omega}
	\newcommand{\ovra}{\overrightarrow}
	\newcommand{\ze}{\zeta}
	\newcommand{\vart}{\vartheta}
	\newcommand{\tri}{\triangle}
	\newcommand{\f}{\frac}
	\newcommand{\iny}{\infty}
	\newcommand{\pro}{\propto}
	\renewcommand{\arraystretch}{1.25}

	\bc {\huge Quantum dynamics in confined pseudo-harmonic oscillator in a time-dependent moving boundary} \\	\vs{1cm} 	{\it Akash Halder \& Amlan K. Roy{\footnote{Email: akroy@iiserkol.ac.in, akroy6k@gmail.com}}\\Department of Chemical Sciences, Indian Institute of Science Education and Research (IISER) Kolkata, Mohanpur-741246, Nadia, WB, India.\\~\\Debraj Nath$^*${\footnote{Email: debrajn@gmail.com ($^*$Corresponding author)}} \\Department of Mathematics, Vivekananda College, Kolkata - 700063, WB, India.} \ec
	\bc {\large {\un{Abstract}}} \ec
	
	In this work, we present analytical solution of Schr\"odinger equation of confined pseudoharmonic potential in presence of a moving boundary condition, for an arbitrary angular momentum state. It turns out that an important quantity to probe the problem is internuclear distance ratio, which depends on the solution of Ermakov equation. The minimum value of time-dependent (TD) Heisenberg uncertainty product  always remains greater than that of the minimum uncertainty product $\hbar/2$. The TD average energy is derived analytically in a closed form and the corresponding average force and average pressure are defined. Moreover, time correlation function of two states for the case of six selected diatomic molecules (CO, NO, ScH, CH, H$_2$, N$_2$) is obtained. It is found to depend on internuclear distance ratio at two different time domains. The TD survival probability and average life-time of molecule in a confined quantum system are defined. Expressions are offered for quantum similarity measure, dissimilarity and quantum similarity index. The latter is given for a pair of molecules. The obtained results are compared with available literature, wherever possible. To our knowledge this is the first detailed report of a non-harmonic central potential in a TD moving boundary condition. \\ ~\\
	\date{\today}


\section{Introduction}
Studies of \emph{confined} quantum systems is of great interest in current research encompassing chemistry, physics and biology. Analysis of systems in sub-domain of $\Gamma$ of space (in contrast to whole space) has much relevance in modeling realistic situations in inhomogeneous media or external fields. Under such extreme conditions, various interesting physical and chemical properties are observed, from both experimental and theoretical perspective. Properties such as electronic structure, chemical reactivity, molecular bond size, energy spectrum, ionization potential, polarizabilities are found to vary in terms of geometrical shape, size and dimension. Some important applications include quantum dot, quantum nanotube, quantum wire, cell model of liquid, molecular sieves, atoms trapped in a nanocavity, polariton, plasmon, gas of bosons etc. Two prototypical confined systems have received maximum attention. They are the quantum harmonic oscillator and hydrogen atom, offering a large number of elegant methods for exact and approximate solution of Schr\"odinger equation (SE) in various dimensions including the N dimension. In both cases, a vast number of references are available; here we cite some representative ones. For the former case, references \cite{campoy2002,sen2006,stevanovic2008,montgomery2010} may be consulted along with a recent review \cite{mukherjee2019}, while for H atom, one may see \cite{burrows2006,ciftci2009,roy2015} and the references therein.

Amongst various problems described by time-dependent Schr\"odinger equation (TDSE), moving boundary condition constitutes an interesting topic. They are commonplace in nature as well in engineering and technological sphere--such as human heart and blood flow (see \cite{canic2020} for a review). A decent number of papers were published in this area in past few decades; in majority cases, \emph{exact} solutions were obtained, mostly for a particle in constant or harmonic oscillator. Also one can find \emph{exact} solution if the boundary moves with a constant speed \cite{yuce2004}. An infinite square well with moving boundary was investigated in \cite{munier1981,cervero1999}. The issue was addressed in the context of an attractive delta potential via a spectral decomposition technique \cite{centeno2020}. The \emph{exact} wave function of a one-dimensional generalized harmonic oscillator with a moving boundary \cite{cervero1999}, was presented in closed form \cite{li2001}, by making use of the algebraic properties of dynamical symmetry and construction of an invariant operator \cite{lejarreta1999}, and also by spectral decomposition \cite{centeno2020}. Apart from these, the H atom in spherical box \cite{rakhmanov2018,centeno2020} and P\"oschl-Teller potential have also been studied \cite{centeno2020}. Besides energy, quantities such as force and pressure have also been considered \cite{rakhmanov2018}. 

Quantum dynamics of a particle in 1D infinite square well having one moving wall is examined \cite{nieto2009}. Controlling a quantum system via a TD boundary condition is proposed \cite{duffin2019}. Recently, average energy, probability density and expectation values have been obtained for TD versions of P\"oschl-Teller potential and its super-symmetric partner (rationally extended P\"oschl-Teller potential) \cite{nath2020} within an oscillating boundary condition. The quantum similarity indices (QSI) in diatomic molecules in presence of a TD generalized P\"oschl-Teller potential has been examined \cite{carbo-dorca2022}. Average energy, average force, time-correlation function, QSI, disequilibrium of TDSE of a particle trapped in a spherical box with a moving boundary wall have been investigated analytically \cite{nath2022}. In another study, Shannon entropy of a confined harmonic oscillator in a TD moving boundary was investigated \cite{nath2023jmc}. Moreover, Heisenberg uncertainty relation, average energy, average force, thermodynamic functions were defined for infinite potential well and trigonometric Rosen-Morse potential in TD quantum system. For high temperature, analytical form of partition function and the corresponding thermodynamic quantities were derived following the Euler-Maclaurin summation formula over a finite as well as an infinite domain for accurate presentation \cite{nath2021,nath2022a}.

Pseudoharmonic oscillator is an important molecular potential \cite{goldman1961,popov2001,ikhdair2007,ikhdair2007a,sever2008,oyewumi2008,oyewumi2012,ikhdair2012,arda2012,akcay2012,yahya2015,rani2018,ghosh2020}, having various applications in chemical physics. Therefore, it has raised considerable interest in molecular quantum chemistry and other field of physics such as quantum mechanics, quantum optics, quantum information. It is used for energy spectrum of linear and nonlinear system. It is exactly solvable for arbitrary angular quantum number $\ell$ in 3 dimension \cite{ikhdair2007,oyewumi2012} and in N dimension \cite{oyewumi2008}. Predominant theoretical methods are Nikiforov-Uvarov (NU) \cite{sever2008}, eigenfunction ansatz \cite{ikhdair2007a,oyewumi2008}, spectrum generating algebra \cite{oyewumi2012}, Laplace transform method \cite{arda2012}, algebraic \cite{akcay2012}, series expansion \cite{rani2018} etc. The Barut-Girardello coherent states have been discussed \cite{popov2001}. The spectrum has been examined in presence of external magnetic field, Aharonov-Bohm field \cite{ikhdair2012}, $\theta$-dependent scalar potential, as well as in the sum of two vector potentials, namely, Dirac magnetic mono-pole and Aharonov-Bohm field \cite{ghosh2020}. Position- and momentum-information theoretic measures (Shannon entropy, Fisher information etc.) in case of diatomic molecules have been studied \cite{yahya2015}. 


In this work, we are interested in the solution of TDSE in presence of pseudoharmonic oscillator potential inside an impenetrable spherical well, with a moving boundary wall. At first we obtain the TD wave function and TD density function. Then we proceed for Heisenberg uncertainty as well as a host of TD quantities such as, average energy, avaerage force, average pressure, time correlation function between two states, auto-correlation function at two times, survival probability and average life time, quantum similarity measure (QSM) between two states, disequilibrium in a state, quantum similarity index (QSI) between two states are analytically derived and visualized through numerically computed results. These results are given for six selected diatomic molecules (CO, NO, ScH, CH, H$_2$ and N$_2$) in arbitrary states represented by quantum numbers $n,\ell,m$.
The article in organized as follows. Section~\ref{sec2.method} provides the required theoretical framework for the problem. The TD nature of internuclear distance is derived in this section. In Sec.~\ref{heisenberg-product}, we consider Heisenberg uncertainty principle for a moving boundary roblem. In Sec.~\ref{average} we present analytical expressions of the necessary TD quantities like average energy, average force, average pressure and discuss their nature. In Sec. \ref{time-co} we express closed analytical form of time-correlation functions (cross-correlation, auto-correlation). Next, the  survival probability, average life time, average probability, disequilibrium as well as QSM and QSI are reported in subsections ~\ref{autocorrelation} and ~\ref{survival}. Representative results are given in the accompanying tables and figures for molecules. Finally, it is concluded with a few remarks in Sec.~\ref{sec5.con}.

\begin{figure}[t]
	\centering
	\includegraphics[width=18cm,height=11cm]{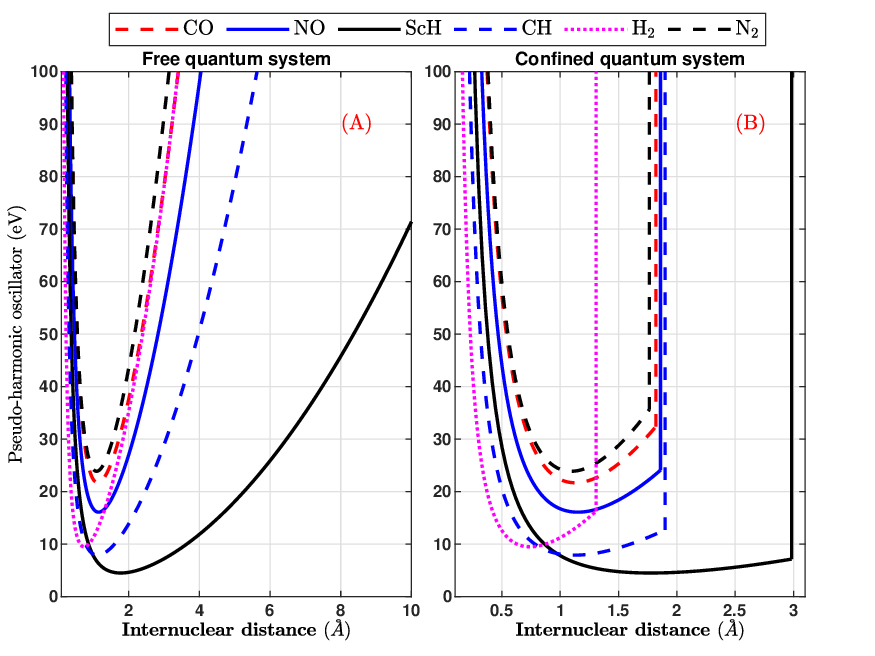}
	\caption{\label{fig1.pot} Pseudoharmonic oscillator potential in a free (A) and a confined (B) molecular systems. For (A), $b=1$ and $t_0=1973.29$ in sec.}
\end{figure}

\section{Methodology}\label{sec2.method}
Let us consider the TDSE for a given molecule, written as follows, 
\begin{equation}\label{eq1.psi}
	\ds\left[-\frac{\hbar^2}{2\mu}\nabla^2+V({\bf r},t)\right]\psi({\bf r},t)=\ds i\hbar\displaystyle\frac{\pad\psi({\bf r},t)}{\pad t}  ,
\end{equation} 
under the influence of a varying TD molecular (pseudoharmonic) potential given below, 
\begin{equation}\label{pot}
\ba{ll}
V({\bf r},t)=\left\{\begin{array}{ll}D_e\left(\frac{r^2}{r_e^2}+\frac{r_e^2}{r^2}\right),&r\le r_0\b(T)  ,  \\
\infty,&r>r_0\b(T) .  \end{array}\right.
\ea 
\end{equation} 
Here $D_e$ and $r_e$ refer to dissociation energy and equilibrium bond length and $r$ is the internuclear distance \cite{atomic-molb}. The latter acts as the radius $r$ of the spherical well with a time dependence, while $\theta$, $\phi$ are independent of time, $t$. To solve this equation, we have invoked a transformation of the form,
\begin{equation}\label{tra2}
	\begin{array}{ll}
	\psi(\mathbf{r},t)=\ds R(r,t)Y_{\ell,m}(\theta,\phi),~~
	r=r_0\b(T)s.
	\end{array} 
\end{equation} 
Here, $\theta, \phi$ vary as $0\le \theta\le \pi$ and $0\le\phi\le 2\pi$ respectively, while $r$ is the TD variable. Let $r_0>0$ and $t_0>0$ be the scale factors of $r$ and $t$, having dimensions of length and time, such that: $t=t_0T$ and $r=r_0\b(T)s$, where $s$ and $T$ are dimensionless variables, whereas $\b(T)$ is a dimensionless scale function of $T$. Here, $\beta (T)$ represents the nature of confinement of the TD Hamiltonian. For pseudoharmonic oscillator potential, Eq.~(\ref{pot}) \cite{ikhdair2007,sever2008,arda2012,oyewumi2012,akcay2012,yahya2015,rani2018,ghosh2020}, the TDSE with moving boundary satisfies the boundary condition: $\psi(0,\theta,\phi,t)=0=\psi(r_0\b(T),\theta,\phi,t)$. Then the TDSE becomes a second-order partial differential equation in terms of dimensionless variables $s$ and $T$ as given below, 
\begin{equation}\label{eq11.xt}
\ba{l}
\ds-\left(\frac{\partial^2 R}{\partial s^2}+\frac{2}{s}\frac{\partial R}{\partial s}\right)+\ds\left[\frac{2\mu D_er_0^4}{\hbar^2r_e^2}\b^4s^2+\frac{L(L+1)}{s^2}\right]R=\ds\left[\frac{2i\mu r_0^2}{\hbar t_0}\b\left(\b\frac{\partial}{\partial T}-\dot{\b}s\frac{\partial}{\partial s}\right)\right]R,
\ea 
\end{equation}
where 
\beq
L(\ell)=-\frac{1}{2}+\sqrt{\left(\ell+\frac{1}{2}\right)^2+\frac{2\mu D_er_e^2}{\hbar^2}},
\eeq 
and $\ell$ denotes the usual angular momentum quantum number. Now, considering the following separation of variables,  
\begin{equation}\label{tra4}
\ba{ll}
R(r)&=\ds e^{\frac{i\alpha(T)s^2}{2}} \frac{X(s)}{s}\tau(T),~~ \a(T)=\ds\frac{\mu r_0^2}{\hbar t_0}\b\dot{\b},~ \tau(T)=\ds\frac{1}{\sqrt{\b^3}}\exp\left[-\ds\frac{it_0\xi^{(c)}}{\hbar}\ds\int^T\frac{dt'}{\b^2(t')}\right],
\ea 
\end{equation}
one obtains,    
\begin{equation}\label{eq.fs}
	\ba{l}
	\ds\frac{d^2 X}{ds^2}+\left[k^2-\lambda^2s^2-\frac{L(L+1)}{s^2}\right]X=0,~
	k=\frac{\sqrt{2\mu r_0^2\xi^{(c)}}}{\hbar},\lam=\frac{\mu r_0^2c_0}{\hbar t_0}.
	\ea 
\end{equation}

\begin{table}[t] 
	\centering
	\begin{tabular}{lllll}\hline\hline
		Sl. & Molecule& \multicolumn{1}{c}{$D_e$} & \multicolumn{1}{c}{$r_e$}&  \multicolumn{1}{c}{$\mu$}\\
		&&\multicolumn{1}{c}{(eV)}&\multicolumn{1}{c}{(\AA)} & \multicolumn{1}{c}{(a.m.u.)}\\\hline
		1&CO&10.845073641&1.1283&6.860586\\
		2&NO&~8.04378&1.1508&6.259494\\
		3&ScH&~2.25& 1.776&0.986040\\
		4&CH&~3.947418665&1.1198&0.929931\\
		5&H$_2$&~4.7446&0.7416&0.50391\\
		6&N$_2$&11.938193&1.0940&7.00335\\\hline\hline
	\end{tabular}
	\centering
	\caption{\label{table1.parameters} The potential parameters for six diatomic molecules, taken from \cite{oyewumi2012}.}
\end{table}

The $T$-dependent factor $\beta$ is obtained from the Ermakov equation \cite{solution.beta}, 
\beq
\beta^3\left[\ddot{\b}+\frac{2D_et_0^2}{\mu r_e^2}\b\right]=c_0^2,
\eeq 
where $c_0$ is a real constant, which is determined later and  $\xi^{(c)}$ is the separation constant. The solution of the Ermakov equation can be written as 
\cite{solution.beta}, 
\beq
\beta(T)=\ds\sqrt{a + b\cos\left(\sqrt{\frac{8D_et_0^2}{\mu r_e^2}}\,T\right)},
\eeq 
where arbitrary constants $a,b$ are functionally related to the constraint,
\beq\label{abc0t0}
a^2=b^2+\frac{\mu r_e^2c_0^2}{2D_et_0^2}.
\eeq 
If $c_0$ is a function of $T$, then $\beta$ can be solved but the method of separation of variables cannot be applied. In this article, we have considered the time independent real variable $c_0$. 
Then functions $X$'s and corresponding eigenenergies are expressed as \cite{kds2007}, 
\begin{equation}
	\ba{l}
	X_{n,\ell}(s)=N_{n,\ell}s^{L+1}e^{-\frac{\lambda}{2}s^2}\,L_n^{(L+\frac{1}{2})}\left[\lambda s^2\right],~
	\xi^{(c)}_{n,\ell}=\frac{\hbar c_0}{t_0}\left(2n+L+\frac{3}{2}\right),~~n=1,2,\cdots, 
	\ea 
\end{equation}
where subscript ``(c)" signifies a quantity under confinement, and $N_{n,\ell}$ is the normalization constant, obtained from the relation, $\ds\int_0^{r_0L}|R_{n,\ell}(r,t)|^2r^2dr=1$, implying that $\ds r_0^3\int_0^{1}|X_{n,\ell}(s)|^2ds=1$. Then one gets,  
\begin{equation}\label{norm}
	N_{n,\ell}=\ds\left[r_0^3\sum\limits_{i=0}^{2n}\frac{B_{2+i,2}\left(c_{n,0},2!c_{n,1},\cdots,(i+1)!c_{n,i}\right)\left[\G\left(i+L+\frac{3}{2}\right)-\G\left(i+L+\frac{3}{2},\lambda\right)\right]}{\lambda^{L+\frac{3}{2}}(2+i)!}\right]^{-\frac{1}{2}},
\end{equation}
where 
\beq
\ba{ll}
c_{n,i}=\ds\left\{\ba{ll}\ds\frac{(-n)_i}{(L(\ell)+\frac{3}{2})_ii!},&~ \mbox{if}~ i\le n\\
0,&~\mbox{if}~ i>n,\ea \right\}.
\ea
\eeq

\begin{figure}[t]
	\centering
	\includegraphics[width=19cm,height=12cm]{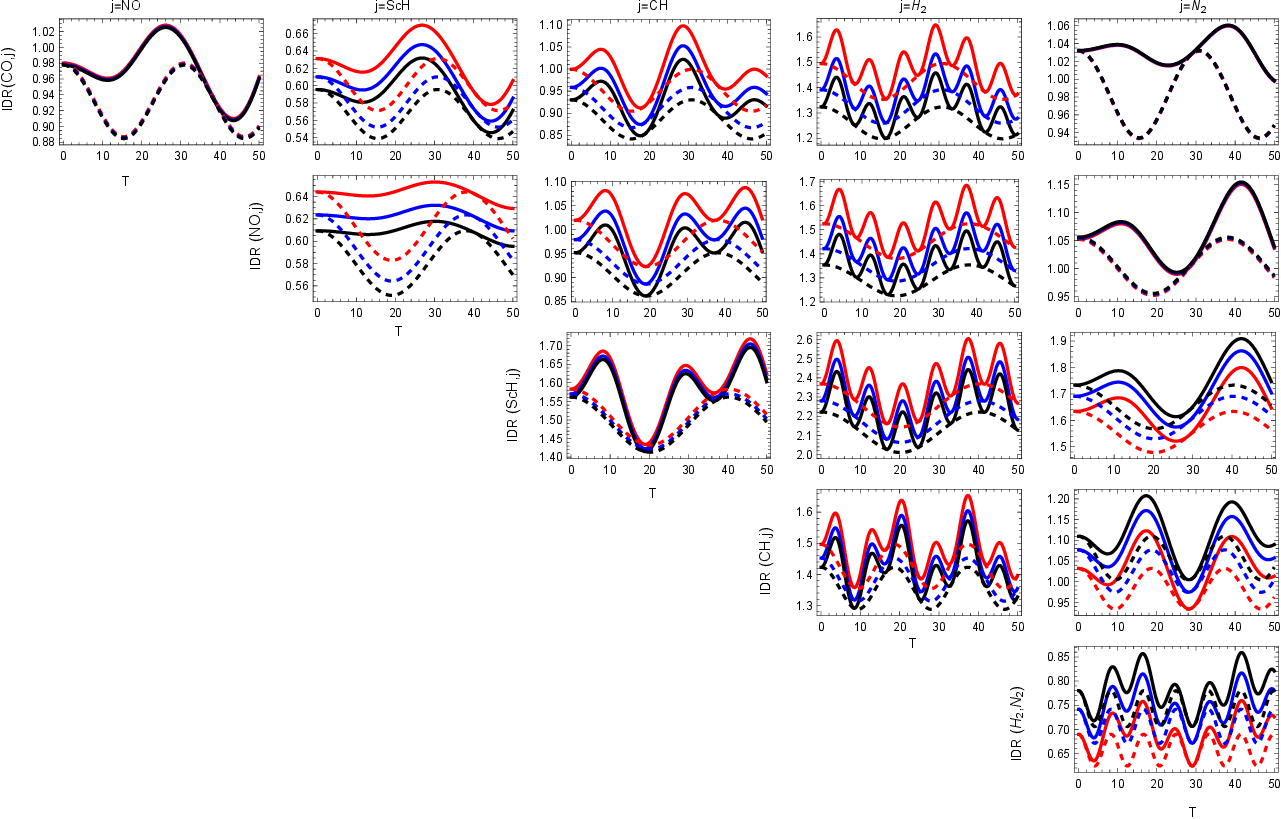}        
	\caption{\label{fig2.radius} The TD internuclear distance ratio of fifteen pairs of diatomic molecules. Dashed and solid lines represent $IDR^{(i,j)}_{n_1,n_2,\ell}(T,0)$ and $IDR^{(i,j)}_{n_1,n_2,\ell}(T,T)$. Red, blue and black lines correspond to $n_1=n_2=1,\ell=0$; $n_1=n_2=2,\ell=1$ and $n_1=n_2=3,\ell=1$ respectively.}
\end{figure}

In the above equation, \((a)_i= a(a+1)...(a+i-1)\) is Pochhammer symbol, \(\Gamma(a)\) signifies Gamma function, \(\Gamma(a,\lam)=\ds\int_{\lam}^{\infty}x^{a-1}e^{-x}dx\) represents the incomplete Gamma function, such that \(\Gamma(a)=\Gamma(a,0)\), while $B_{m,\ell}(x_1,x_2,\dots,x_{m-l+1})$ denotes Bell polynomial \cite{riordan1980}, defined as, 
\begin{equation*}
	B_{m,\ell}(x_1,x_2,\dots,x_{m-l+1})= \ds\sum\limits_{\widehat{\pi}(m,\ell)}\frac{m!}{j_1!j_2!\dots j_{m-\ell+1}!} 
	\left(\frac{x_1}{1!}\right)^{j_1} \left(\frac{x_2}{2!}\right)^{j_2} \dots  \bigg( \frac{x_{m-\ell+1}}{(m-\ell+1)!} \bigg)^{j_{m-\ell+1}}.
\end{equation*} 
Here, $\widehat{\pi}(m,\ell)$ refers to the set of partitions, such that, $\ds j_1+j_2+\dots+j_{m-\ell+1}=\ell,~j_1+2j_2+\dots+(m-\ell+1)j_{m-\ell+1}=m,$ where $c_0,r_0$ are real constants which satisfy the following constraints: 
${}_1F_1\left[-n,L+\frac{3}{2},\frac{\mu c_0r_0^2}{\hbar t_0}\right]=0,~
{}_1F_1\left[-n+1,L+\frac{5}{2},\frac{\mu c_0r_0^2}{\hbar t_0}\right]\ne 0$.
If one chooses $c_0=\ds\sqrt{\frac{2t_0^2D_e}{\mu r_e^2}}$, then the eigenvalues, $\xi^{(c)}_{n,\ell}$ can be expressed \cite{ghosh2020,oyewumi2012} as: 
$\xi^{(c)}_{n,\ell}=\ds\sqrt{\frac{2\hbar^2D_e}{\mu r_e^2}}\left(2n+L(\ell)+\frac{3}{2}\right)$. 

In this case, one can find the TD internuclear distance as, $r_{n,\ell}=\ds\sqrt{\frac{\hbar t_0\lambda_{n,\ell}}{\mu c_0}}\,\b(T)$, where $\lambda_{n,\ell}$ is the largest root of ${}_1F_1\left[-n,L+\frac{3}{2},\lambda_{n,\ell}\right]=0$. In particular, for $n=1,2,3,...$, one obtains, 
\beq
\left.\ba{lll}
r_{n,\ell}&=\ds\sqrt{\frac{\hbar t_0}{\mu c_0}\left(L(\ell)+\frac{3}{2}\right)}\,\b(T),&n=1\\
&=\ds\sqrt{\frac{\hbar t_0}{\mu c_0}\left(L(\ell)+\frac{5}{2}+\sqrt{L(\ell)+\frac{5}{2}}\right)}\,\b(T),&n=2\\
&=\ds\sqrt{\frac{\hbar t_0}{\mu c_0}\left(L(\ell)+\frac{7}{2}+2\sqrt{L(\ell)+\frac{7}{2}}\cos\left[\frac{1}{3}\tan^{-1}\sqrt{L(\ell)+\frac{5}{2}}\right]\right)}\,\b(T),~~&n=3.
\ea\right\}.
\eeq 
Therefore, the complete solution of TDSE can be finally written as,
\beq
\ba{lr}
\psi_{n,\ell,m}(r,\theta,\phi,T)=\frac{N_{n,\ell}\Theta_{\ell,m}(\theta)}{\sqrt{2\pi[\b(T)]^3}}s^{L}\,{}_1F_1\left[-n,L+\frac{3}{2},\lambda s^2\right]\exp\left\{-\frac{\lambda\,s^2}{2}+i\left(m\phi+\frac{\alpha(T) s^2}{2}-\frac{\sqrt{\mu} r_e\xi_{n,\ell}^{(c)}\tan^{-1}\left[\sqrt{\frac{a-b}{a+b}}\tan\left(\sqrt{\frac{2D_et_0^2}{\mu r_e^2}} T\right)\right]}{\sqrt{2D_e(a^2-b^2)\hbar^2}}\right)\right\},
\ea
\eeq
where $s=\ds\frac{r}{r_{0,(n,\ell)}\b(T)}$, $n=1,2,3,\cdots;$ $\ell=0,1,\cdots,n-1$ and $m=0,\pm1,\pm2,\cdots,\pm\ell$. For the confined quantum system $n=1,2,3,\cdots$ represent the ground, first, second excited states and so on. It may be noted that, when $t_0,c_0\rightarrow\infty$, $\beta=\left(\frac{\mu r_e^2c_0^2}{2D_et_0^2}\right)^{\frac{1}{4}}.$

\begin{table}[t] 
	\centering
	\scalebox{.9}{\begin{tabular}{llllllllllllll}\hline\hline
			$n$& $\ell$ & \multicolumn{2}{c}{CO} & \multicolumn{2}{c}{NO}&  \multicolumn{2}{c}{ScH}& \multicolumn{2}{c}{CH}& \multicolumn{2}{c}{H$_2$}& \multicolumn{2}{c}{N$_2$}\\\hline
			&&\multicolumn{1}{c}{$r_*$(\AA)}&\multicolumn{1}{c}{$r^*$(\AA)} &\multicolumn{1}{c}{$r_*$(\AA)}&\multicolumn{1}{c}{$r^*$(\AA)}&\multicolumn{1}{c}{$r_*$(\AA)}&\multicolumn{1}{c}{$r^*$(\AA)}&\multicolumn{1}{c}{$r_*$(\AA)}&\multicolumn{1}{c}{$r^*$(\AA)}&\multicolumn{1}{c}{$r_*$(\AA)}&\multicolumn{1}{c}{$r^*$(\AA)}&\multicolumn{1}{c}{$r_*$(\AA)}&\multicolumn{1}{c}{$r^*$(\AA)}\\\hline
			\multicolumn{14}{c}{b=0}\\\hline
			1&0&1.13095&1.13095&1.15375&1.15375&1.79131&1.79131&1.1317&1.1317&0.756311&0.756311&1.0965&1.0965\\
			2&0&1.17162&1.17162&1.19717&1.19717&1.91966&1.91966&1.22245&1.22245&0.841459&0.841459&1.13537&1.13537\\
			2&1&1.17163&1.17163&1.19719&1.19719&1.91992&1.91992&1.2227&1.2227&0.842027&0.842027&1.13538&1.13538\\
			3&0&1.20214&1.20214&1.2298&1.2298&2.01791&2.01791&1.2922&1.2922&0.90779&0.90779&1.16453&1.16453\\
			3&1&1.20216&1.20216&1.22981&1.22981&2.01817&2.01817&1.29245&1.29245&0.908349&0.908349&1.16454&1.16454\\
			3&2&1.20218&1.20218&1.22984&1.22984&2.01869&2.01869&1.29294&1.29294&0.909464&0.909464&1.16457&1.16457\\\hline
			\multicolumn{14}{c}{b=0.1}\\\hline
			1&0&1.07588&1.18883&1.09757&1.2128&1.70409&1.883&1.0766&1.18962&0.719485&0.795022&1.04311&1.15262\\
			2&0&1.11457&1.23159&1.13888&1.25845&1.82619&2.01791&1.16292&1.28502&0.800487&0.884528&1.08008&1.19348\\
			2&1&1.11458&1.2316&1.13889&1.25846&1.82644&2.01819&1.16316&1.28528&0.801028&0.885126&1.0801&1.19349\\
			3&0&1.14361&1.26367&1.16992&1.29275&1.91966&2.1212&1.22928&1.35834&0.863588&0.954254&1.10783&1.22414\\
			3&1&1.14362&1.26369&1.16993&1.29276&1.9199&2.12147&1.22952&1.3586&0.86412&0.954842&1.10784&1.22415\\
			3&2&1.14364&1.26371&1.16996&1.29279&1.9204&2.12201&1.22999&1.35912&0.865181&0.956014&1.10786&1.22417\\\hline\hline
	\end{tabular}}
	\centering\caption{\label{table2.radius} Calculated short $(r_*)$ and long $(r^*)$ range of internuclear distance, for $t_0=$1973.29 sec.}
\end{table}

Therefore, the general solution of the moving boundary problem is expressed as a superposition of all possible single mode states 
\beq
\Psi(r,\theta, \phi, t)=\ds\sum\limits_{n=1}^{\infty}\sum\limits_{\ell=0}^{n-1}\sum\limits_{m=-\ell}^{\ell} c_{n,\ell.m} \psi_{n,\ell,m}(r,\theta,\phi,t),
\eeq
where the time-independent coefficients $c_{n,\ell,m}$ are obtained from the following relation
\beq
c_{n,\ell,m}=\int_{0}^{r_0 \b} r^2 dr \int d\Omega \ds \psi^{*}_{n,\ell,m}(\mathbf{r},0) \Psi(\mathbf{r},0),~d\Omega=\sin \theta d\theta d\phi.
\eeq
For numerical calculations, following conversions are adopted: $1\, amu = 931.494028\, MeV/c^2$, $1\, cm^{-1} = 1.239841875 \times 10^{-4}\,eV$ and $c\hbar =1973.29\, eV\AA$, where $c$ denotes the speed of light. The pseudoharmonic oscillator potential is plotted in Fig.~\ref{fig1.pot} in free and confined quantum systems, for six diatomic molecules, \emph{viz.}, CO, NO, ScH, CH, H$_2$ and N$_2$ in panels (A) and (B) respectively. The set of parameters in Table \ref{table1.parameters}, are adopted from \cite{oyewumi2012}. In the following section, we  investigate the nature of solutions in these molecules, with respect to various TD properties. 
To proceed further, at first, we present analytical results of TD density, TD internuclear distance. Representative data and figures are given for 6 molecules mentioned earlier. 

\subsection{The TD density function and internuclear distance}
The density function of a given state $\psi_{n,\ell,m}({\bf r}, t)$ is defined by $\rho_{n,\ell,m}({\bf r},t)=d_{n,\ell}(r,t)\Om_{\ell,m}(\theta,\phi)$ where: 
\beq\label{den.con}
d^{(c)}_{n,\ell}(r,t)=\ds\left\{\ba{ll}\ds\frac{N_{n,\ell}^2}{\b^3}\left(\frac{r}{r_0\b}\right)^{2L}e^{-\frac{\lam r^2}{r_0^2\b^2}}\,\left({}_1F_1\left[-n,L+\frac{3}{2},\frac{\lambda r^2}{r_0^2 \b^2}\right]\right)^2, &0\le r\le r_0\b(T)\\0,&r>r_0\b(T)\ea \right.
\eeq
and 
\beq
\Om_{\ell,m}(\theta,\phi)=\ds\frac{\left[\Theta_{\ell,m}(\theta)\right]^2}{2\pi}, \ 0\le\theta\le \pi, 0\le\phi\le 2\pi
\eeq
are respectively radial and angular density functions.
	
For a confined quantum system, TD effective domain of the radial wave function at a time $t$ is \emph{exactly} defined by $[0,r_{0,(n,\ell)}\b(T)]$, for a given state $\psi_{n,\ell,m}(r,\theta,\phi,t)$. This signifies a movement in a confined environment under a moving boundary condition with a TD radius $r_0\b(T)$, which lies in a finite region $\left[r_0\sqrt{a-b},~ r_0\sqrt{a+b}\right]$. Then the short and long range of internuclear distances are denoted as, $[0,r_*]$ and $[0,r^*]$, with $r_*, r^*$ being given by, 
\beq
r_*=r_{0,(n,\ell)}\sqrt{\sqrt{b^2+\frac{\mu r_e^2c_0^2}{2D_et_0^2}}-b},~ r^*=r_{0,(n,\ell)}\sqrt{\sqrt{b^2+\frac{\mu r_e^2c_0^2}{2D_et_0^2}}+b}. 
\eeq

Here $b$ is a free parameter and $r_{0,(n,\ell)}$ depends on $n,\ell$ quantum numbers. For the special case of $n=1,2,3$, one can write the followings, 
\beq
\left.\ba{lll}
r_{0,(n,\ell)}&=\sqrt{\frac{\hbar t_0}{\mu c_0}\left(L(\ell)+\frac{3}{2}\right)},&n=1\\
&=\sqrt{\frac{\hbar t_0}{\mu c_0}\left(L(\ell)+\frac{5}{2}+\sqrt{L(\ell)+\frac{5}{2}}\right)},&n=2\\
&=\sqrt{\frac{\hbar t_0}{\mu c_0}\left(L(\ell)+\frac{7}{2}+2\sqrt{L(\ell)+\frac{7}{2}}\cos\left[\frac{1}{3}\tan^{-1}\sqrt{L(\ell)+\frac{5}{2}}\right]\right)},&n=3
\ea\right\}. 
\eeq 
It may be noted that, when $b=0$, $r_*=r^*=\ds r_{0,(n,\ell)}\left(\frac{\mu r_e^2c_0^2}{2D_et_0^2}\right)^{\frac{1}{4}}$, but $r_*\rightarrow 0$ and $r^*\rightarrow\infty$ if $b\rightarrow\infty$. Practically, $0<r_*<r^*<\infty$, and both of them are non-zero finite, and therefore, $b$ has reasonable finite values. If $b=0$, then the confined system is time independent and the corresponding boundary wall is fixed. In this case, internuclear distance of a molecule is fixed for a particular quantum number $n,\ell$. A very important quantity for our future discussion is given by, internuclear distance ratio (IDR) $IDR^{(i,j)}_{n_1,n_2,\ell}(T_1,T_2)=\ds\frac{r_{0,(n_1,\ell)}^{(i)}\beta^{(i)}(T_1)}{r_{0,(n_2,\ell)}^{(j)}\beta^{(j)}(T_2)}$, which is plotted in Fig.~\ref{fig2.radius}. It is observed that IDR fluctuates with respect to $T$. But $IDR^{(i,j)}_{n_1,n_2,\ell}(T_1,T_2)$ is a periodic function of $T_1$ for fixed $T_2$ and \emph{vice versa}. 

\begin{figure}[t]
	\centering
	\includegraphics[width=6.5cm,height=6cm]{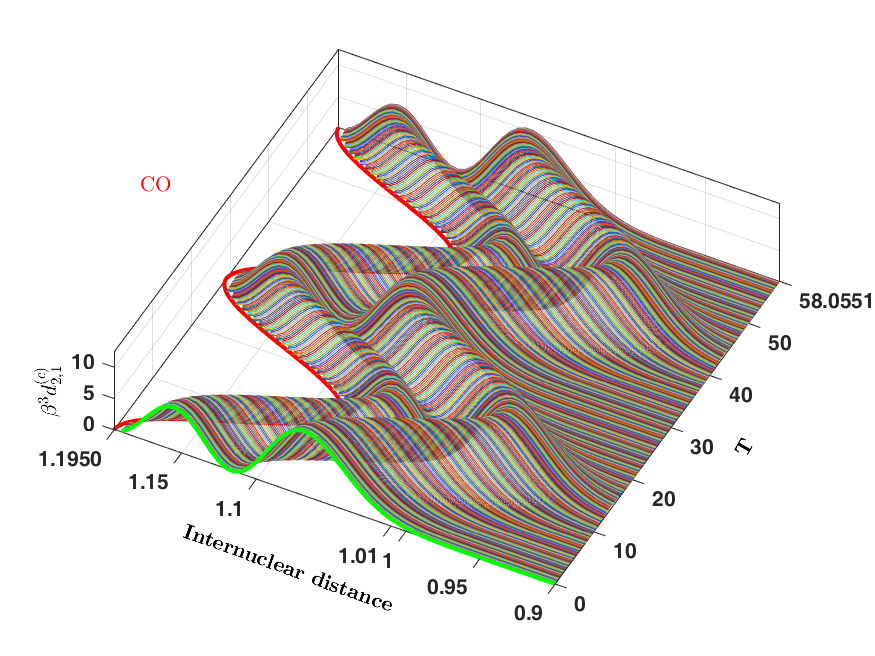}~\includegraphics[width=6.5cm,height=6cm]{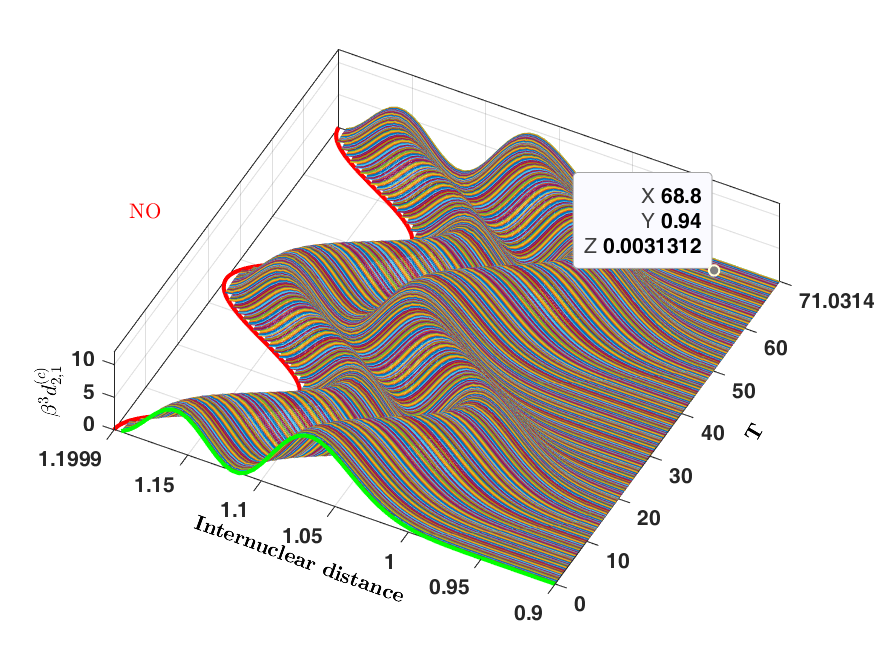}~\includegraphics[width=6.5cm,height=6cm]{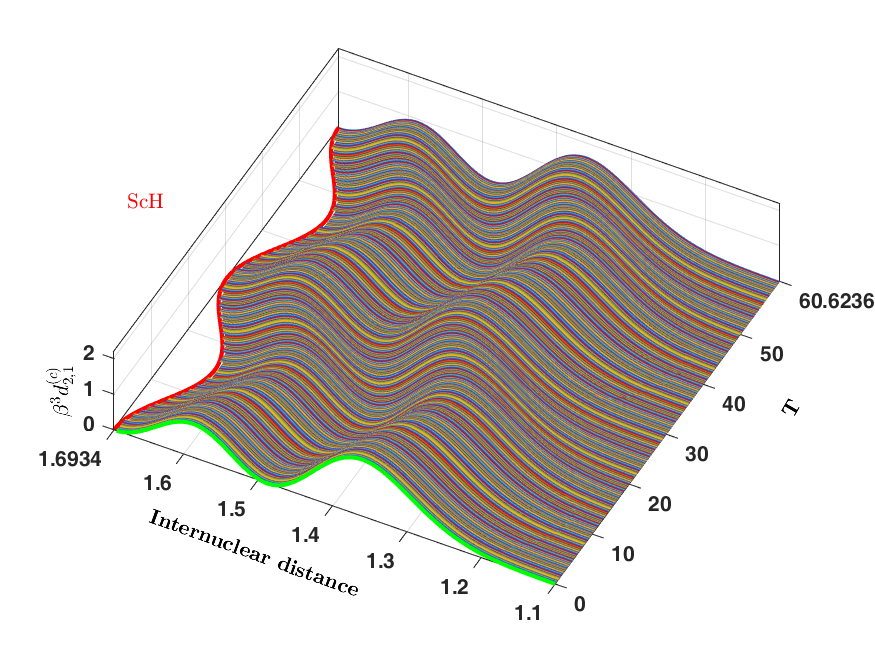}\\
	\includegraphics[width=6.5cm,height=6cm]{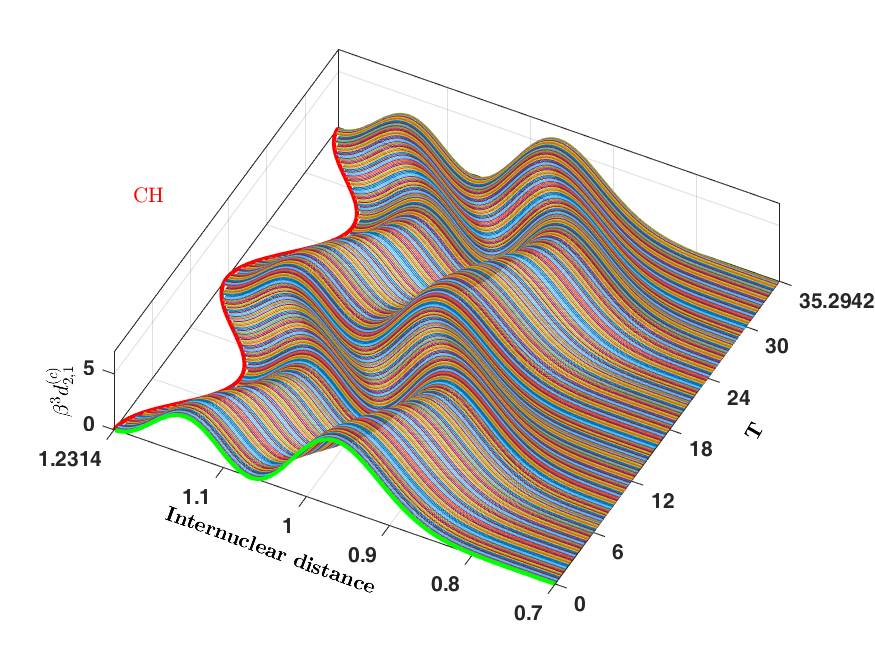}~\includegraphics[width=6.5cm,height=6cm]{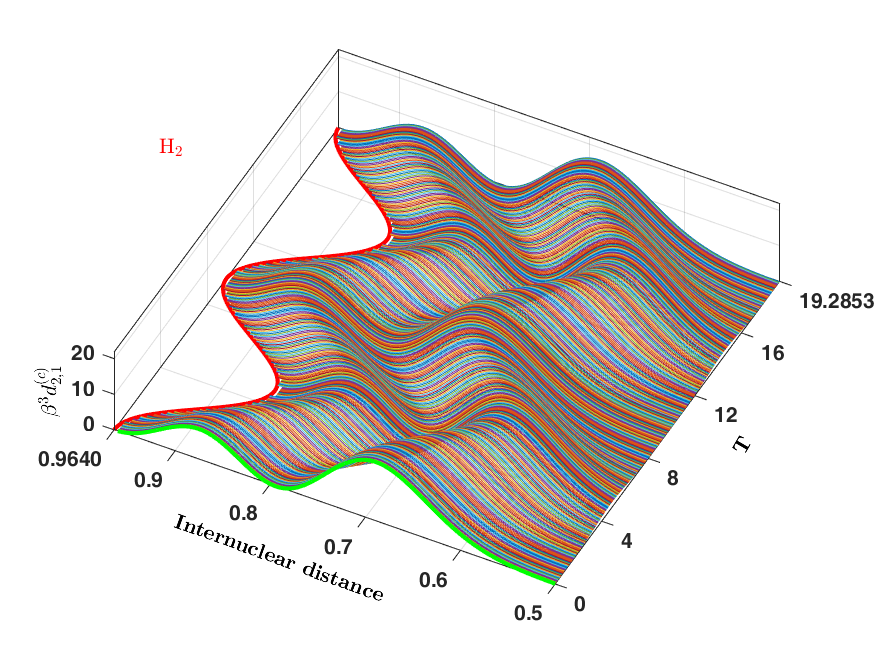}~\includegraphics[width=6.5cm,height=6cm]{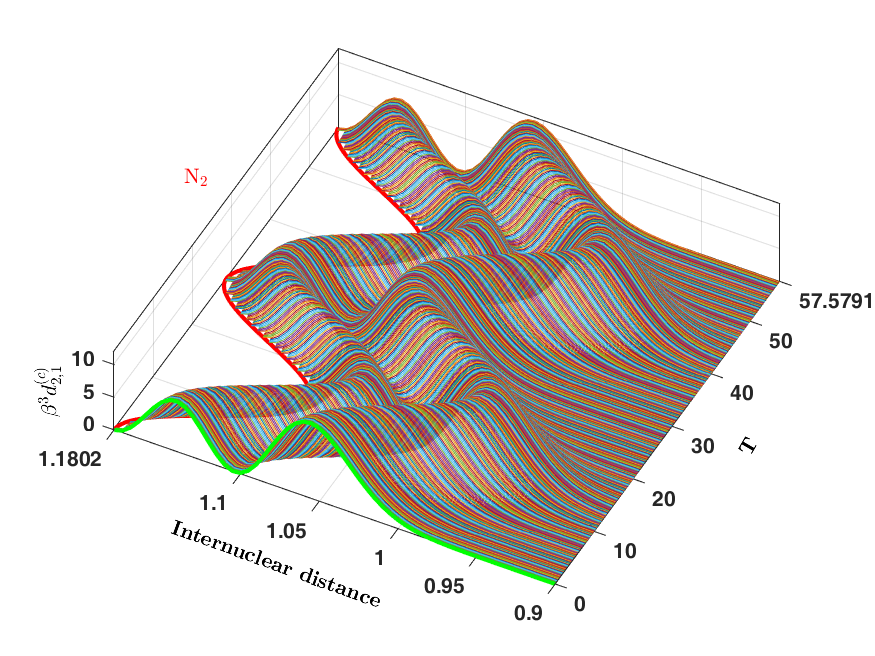}
	\caption{\label{fig2.density} Plot of TD radial density functions of first excited states of six diatomic molecules for $b=1$ and $t_0=$1973.29 sec.}
\end{figure}

Now, the radial density function, $d_{n,\ell}^{(c)}(r,t)$, given in Eq.~(\ref{den.con}), is visualized for six molecules in terms of $\beta^3d_{n,\ell}^{(c)}(r,t)$, in Fig.~\ref{fig2.density} for $n=2,\ell=1$, $b=0.1, t_0=1973.29$. It is clear that the boundary wall is impenetrable. The initial solution and moving boundary conditions are depicted by green and red lines. The molecules are provided in the respective panels. One also sees that the internuclear distance satisfies the inequality $0<r_*<r^*<\infty$. The estimated values of shorter $(r_*)$ and longer $(r^*)$ range of internuclear distance are tabulated in Table~\ref{table2.radius}, for $t_0=1973.29\, sec$ and $b = 1$, for some representative states, having quantum numbers $n=1,2,3$ and $\ell=1$. 


\section{Heisenberg uncertainty}\label{heisenberg-product}
The expectation of $r^j$ in an arbitrary state $\psi_{n,\ell,m}({\bf r},t)$ can be written as:
\begin{equation}
	\begin{array}{ll}\label{un.r.i}		\left\langle r^j\right\rangle_{n,\ell,m}&=\ds\int \left[\psi_{n,\ell,m}({\bf r},t)\right]^* r^j \psi_{n,\ell,m}({\bf r},t)\,d{\bf r}
		=(r_{0,(n,\ell)} \beta)^j \left\langle s^j\right\rangle_{n,\ell,m}, ~d{\bf r}=\int_0^{r_{0,(n,\ell)}\b}drr^2\int d\Om, 
	\end{array} 
\end{equation}
where 
\beq\label{ex.sj}
\left\langle s^j\right\rangle_{n,\ell,m}=\ds r_{0,(n,\ell)}^3N_{n,\ell}^2\sum\limits_{i=0}^{2n}\frac{B_{2+i,2}\left(c_{n,0},2!c_{n,1},\cdots,(i+1)!c_{n,i}\right)\left[\G\left(i+L+\frac{j+3}{2}\right)-\G\left(i+L+\frac{j+3}{2},\lambda\right)\right]}{\lambda^{L+\frac{j+3}{2}}(2+i)!}.
\eeq 
The analytical form of $\left\langle s^j\right\rangle_{n,\ell,m}$ is defined for $L+\frac{j+3}{2}>0$. Hence we can find $\left\langle s^j\right\rangle_{n,\ell,m}$ for $j=-3,2,1,0,1,2,3,\cdots$. Thus, 
\begin{equation}
\begin{array}{ll}\label{un.r}
	\left\langle r\right\rangle_{n,\ell,m}&=\ds\int \left[\psi_{n,\ell,m}({\bf r},t)\right]^* r \psi_{n,\ell,m}({\bf r},t)\,d{\bf r}
	=r_{0,(n,\ell)} \beta \left\langle s\right\rangle_{n,\ell,m}\\
		\left\langle r^2\right\rangle_{n,\ell,m}&=\ds\int \left[\psi_{n,\ell,m}({\bf r},t)\right]^* r \psi_{n,\ell,m}({\bf r},t)\,d{\bf r}
	=r_{0,(n,\ell)}^2 \beta^2 \left\langle s^2\right\rangle_{n,\ell,m}
	
\end{array} 
\end{equation}
Therefore, root mean square (RMS) of $r$ can be given as below:  
\beq
(\Delta r)_{n,\ell,m}= \sqrt{\left\langle r^2\right\rangle_{n,\ell,m}-\left\langle r\right\rangle_{n,\ell,m}^2}=r_{0,(n,\ell)} \beta (\Delta s)_{n,\ell,m}, 
\eeq
where $(\Delta s)_{n,\ell,m}=\sqrt{\left\langle s^2\right\rangle_{n,\ell,m}-\left\langle s\right\rangle_{n,\ell,m}^2}$. The term $r_{0,(n,\ell)}\b$ is the scale factor of RMS of $r$.
The expectation of radial momentum operator, $p_r=-i\hbar \left(\frac{\partial }{\partial r}+\frac{1}{r}\right)$, is given as, 
\beq
\left\langle p_r \right\rangle_{n,\ell,m}=\frac{\alpha \hbar}{r_{0,(n,\ell)} \beta} \left\langle s\right\rangle_{n,\ell,m}.
\eeq

\begin{table}[t] 
	\centering
	\begin{tabular}{ccrrrrrr}\hline\hline
		& & \multicolumn{6}{c}{$\min{[(\Delta r)_{n,\ell,m} (\Delta p)_{n,\ell,m}]/\hbar}$} \\\hline
		$n$&$\ell$&\multicolumn{1}{c}{CO} &\multicolumn{1}{c}{NO}  &\multicolumn{1}{c}{ScH}&\multicolumn{1}{c}{CH}&\multicolumn{1}{c}{H$_2$}&\multicolumn{1}{c}{N$_2$}\\\hline
		1&0&0.013012&0.013569&0.024017&0.026408&0.034733&0.012841\\
		2&0&0.062633&0.065285&0.114661&0.125859&0.164591&0.061820\\
		2&1&0.062632&0.065285&0.114645&0.125834&0.164483&0.061819\\
		3&0&0.134269&0.139830&0.241231&0.263621&0.339094&0.132561\\
		3&1&0.134267&0.139828&0.241201&0.263571&0.338888&0.132559\\
		3&2&0.134264&0.139825&0.241139&0.263471&0.338478&0.132557\\\hline\hline
	\end{tabular}
	\centering
	\caption{\label{table3.uncertainty} Minimum value of uncertainty product for six molecules in a TD moving boundary problem.}
\end{table}

 Due to the presence of position as well as TD phase $e^{\frac{i \alpha s^2}{2}}$ in radial wave function, the expectation $\left\langle p_r \right\rangle_{n,\ell,m}$ in a  moving boundary in nonzero $\left\langle p_r \right\rangle_{n,\ell,m}\ne0$. But for free boundary or time-independent fixed boundary it is zero. Thus, $ \left\langle p_r^2 \right\rangle_{n,\ell,m}$ is expressed as 
 \beq
 \left\langle p_r^2 \right\rangle_{n,\ell,m}=\frac{ \hbar^2}{r_{0,(n,\ell)}^2 \beta^2} \left[\a^2\left\langle s^2\right\rangle_{n,\ell,m} +\mathcal{J}_{n,\ell}-\left\langle\frac{1}{s}\right\rangle_{n,\ell,m}\right],
 \eeq
 where
 \beq
 \ba{ll}
 \mathcal{J}_{n,\ell}&=\ds r_{0,(n,\ell)}^3N_{n,\ell}^2\int_0^1 \left[\frac{dX_{n,\ell}(s)}{ds}\right]^2ds\\
 &=\{(L+1)^2-2(L+1)\left(L+\frac{1}{2}\right)\}\left\langle \frac{1}{s^2}\right\rangle_{n,\ell,m}-\lambda^2\left\langle s^2\right\rangle_{n,\ell,m}+2 \lambda (L+\frac{3}{2})\\
 &\quad + \ds\frac{4 \lambda^2 n^2r_{0,(n,\ell)}^3 N_{n,\ell}^2}{\left(L+\frac{3}{2}\right)^2} \sum\limits_{i=0}^{2n-2}\frac{B_{2+i,2}\left(c_{n-1,0},2!c_{n-1,1},\cdots,(i+1)!c_{n-1,i}\right)\left[\G\left(i+L+\frac{5}{2}\right)-\G\left(i+L+\frac{5}{2},\lambda\right)\right]}{\lambda^{L+\frac{5}{2}}(2+i)!},
 \ea 
 \eeq
 \beq
 \ba{ll}
 c_{n-1,i}=\ds\left\{\ba{ll}\ds\frac{(-n+1)_i}{(L(\ell)+\frac{5}{2})_ii!},&~ \mbox{if}~ i\le n-1\\
 0,&~\mbox{if}~ i>n-1,\ea \right\}.
 \ea
 \eeq
 Therefore, RMS of $\left\langle p_r \right\rangle_{n,\ell,m}$ can be written as follows:  
 \beq
 (\Delta p_r)_{n,\ell,m}= \frac{\hbar}{r_{0,(n,\ell)} \beta}\sqrt{\a^2\left(\Delta s\right)_{n,\ell,m}^2 +\mathcal{J}_{n,\ell}-\left\langle\frac{1}{s}\right\rangle _{n,\ell,m}}.
 \eeq
 It then follows that, the Heisenberg uncertainty product in a given state is written as:  
 \beq
 (\Delta r)_{n,\ell} (\Delta p_r)_{n,\ell}= \hbar\sqrt{\a^2(\Delta s)_{n,\ell}^4+\left(\mathcal{J}_{n,\ell}-\left\langle\frac{1}{s}\right\rangle_{n,\ell} \right)(\Delta s)_{n,\ell}^2}\ge \hbar(\Delta s)_{n,\ell}\sqrt{\left(\mathcal{J}_{n,\ell}-\left\langle\frac{1}{s}\right\rangle_{n,\ell} \right)} .
 \eeq
 Heisenberg uncertainty product of position and momentum is an important quantity. 
 It is independent of azimuthal quantum number $m$. The minimum value of uncertainty product due to moving boundary condition in a confined quantum system  is equal to $\hbar\sqrt{\mathcal{J}_{n,\ell} -\left\langle 1/s\right\rangle_{n,\ell,m}}$. 
 The uncertainty product is a periodic function of time and it oscillates between 
 $\hbar(\Delta s)_{n,\ell}\sqrt{\mathcal{J}_{n,\ell}-\left\langle 1/s\right\rangle_{n,\ell,m}}$ and $\hbar\sqrt{\frac{\mu^2 r_{0,(n,\ell)}^4}{\hbar^2 t_0^2}\frac{2D_et_0^2b^2}{\mu r_e^2}(\Delta s)_{n,\ell,m}^4+\left(\mathcal{J}_{n,\ell}-\left\langle\frac{1}{s}\right\rangle_{n,\ell}\right)(\Delta s)_{n,\ell}^2}$. 
  
 \begin{figure}[t]
 	\centering
 	\includegraphics[width=19cm,height=11cm]{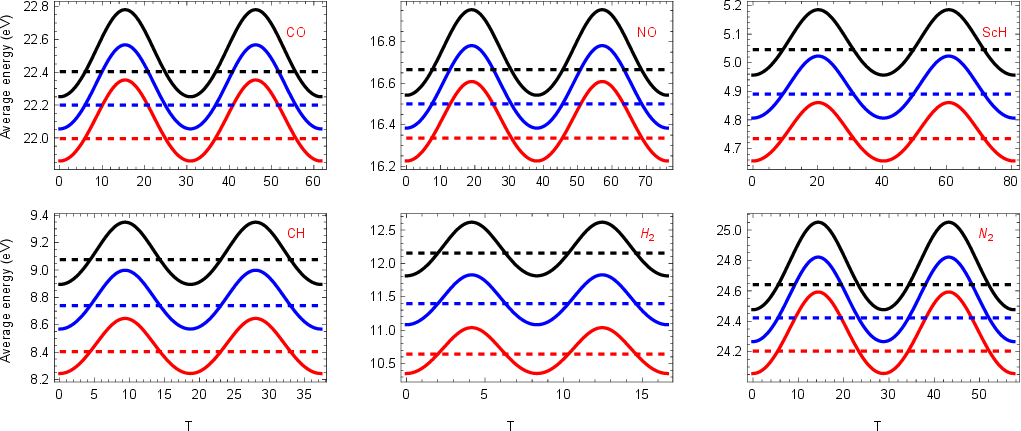}
 	\caption{\label{fig3.average-energy} Plot of the TD average energy (in eV) of molecules, for $b=1$ and $t_0=1973.29\,$sec. Solid and dashed lines represent $\xi^{(c)}_{n,\ell}$ and average energy. Red, blue and black lines are drawn for $n=1,\ell=0$; $n=2,\ell=1$; and $n=3,\ell=1$ states respectively}.
 \end{figure}
 
 \section{Time dependent average energy, average force and average pressure}\label{average}
The average energy of a TD single mode normalized state $\psi_{n\ell,m}({\bf r},t)$ is defined by \cite{rakhmanov2018,nath2020,nath2021,carbo-dorca2022,nath2022,nath2022a,nath2022_nld,nakamura2011}, 
\begin{equation}
\begin{array}{ll}\label{average.en}
\left\langle E\right\rangle_{n,\ell,m}&=\ds i\hbar\int \left[\psi_{n,\ell,m}({\bf r},t)\right]^*\frac{\partial \psi_{n,\ell,m}({\bf r},t)}{\partial t}\,d{\bf r}
=\ds \frac{\xi^{(c)}_{n,\ell}}{\b^2}+\ds\frac{\mu r_{0,(n,\ell)}^2}{2t_0^2}\left(\left[\dot{\b}\right]^2-\b\ddot{\b}\right)\left\langle s^2\right\rangle_{n,\ell}.
\end{array} 
\end{equation}
Due to the degeneracy present, $\left\langle E\right\rangle_{n,\ell,m}$ is independent of $m$. One can find the analytical expression of this expectation value, $\left\langle s^2\right\rangle_{n,\ell}$, from Eq.~(\ref{ex.sj}). The average energy at $t=0$ is: $\left\langle E(0)\right\rangle_{n,\ell,m}=\ds \frac{\xi^{(c)}_{n,\ell}}{a+b}+\frac{2bD_er_{0,(n,\ell)}^2}{r_e^2}\left\langle s^2\right\rangle_{n,\ell}$. It may be noted that if $b=0$, then $\left\langle E\right\rangle_{n,\ell,m}=\frac{\xi^{(c)}_{n,\ell}}{\b^2}$. 

For  the special choice of $\frac{c_0}{t_0}=\ds\sqrt{\frac{2D_e}{\mu r_e^2}}$, one can write $\left\langle E\right\rangle_{n,\ell,m}=\xi^{(c)}_{n,\ell}$, which is equal to the ro-vibrational energy of a molecule in confined quantum system with pseudoharmonic oscillator potential in a spherical box of finite radius, while $\left\langle E\right\rangle_{n_1,\ell_1,m_1}-\left\langle E\right\rangle_{n_2,\ell_2,m_2}=\xi^{(c)}_{n_1,\ell_1}-\xi^{(c)}_{n_2,\ell_2}$. In \textit{free quantum system}, the ro-vibrational energy of a molecule in pseudoharmonic potential is obtained as \cite{ghosh2020,oyewumi2012} (superscript ``(f)" refers a quantity in the \emph{free} system), 
\beq
\xi^{(f)}_{n,\ell}=\ds\hbar\om^{(f)}\left(2n+L(\ell)-\frac{1}{2}\right),~n=1,2,\cdots,~\ell=0,1,\cdots,n-1,~m=0,\pm1,\pm2,\cdots .
\eeq 

\begin{table}[h]
	\centering
	\scalebox{.9}{\begin{tabular}{llllllllllllll}\hline\hline
			$n$&$\ell$ & \multicolumn{2}{c}{CO} & \multicolumn{2}{c}{NO}&  \multicolumn{2}{c}{ScH}& \multicolumn{2}{c}{CH}& \multicolumn{2}{c}{H$_2$}& \multicolumn{2}{c}{N$_2$}\\ \hline
			&&\multicolumn{1}{c}{Min}&\multicolumn{1}{c}{Max} &\multicolumn{1}{c}{Min}&\multicolumn{1}{c}{Max}&\multicolumn{1}{c}{Min}&\multicolumn{1}{c}{Max}&\multicolumn{1}{c}{Min}&\multicolumn{1}{c}{Max}&\multicolumn{1}{c}{Min}&\multicolumn{1}{c}{Max}&\multicolumn{1}{c}{Min}&\multicolumn{1}{c}{Max}\\\hline
			1&0&21.8589&22.3522&16.2249&16.6078&4.65494&4.85924&8.24094&8.64281 & 10.3362&11.022&24.0569&24.5922\\
			2&0&22.0547&22.5661&16.3834&16.7808&4.80503&5.02178&8.56568&8.99435 & 11.0663&11.8127&24.2665&24.8213\\
			2&1&22.0552&22.5666&16.3838&16.7812&4.80637&5.02315&8.56924&8.99799 & 11.0813&11.828&24.267&24.8218\\
			3&0&22.2502&22.7801&16.5418&16.9539&4.95545&5.184&8.89123&9.34508 & 11.799&12.6009&24.4758&25.0506\\
			3&1&22.2507&22.7806&16.5422&16.9543&4.95678&5.18537&8.89479&9.34872 & 11.8139&12.6162&24.4763&25.0511\\
			3&2&22.2516&22.7816&16.5431&16.9552&4.95945&5.18809&8.90191&9.356 & 11.8437&12.6469&24.4773&25.0521\\\hline
			\hline
	\end{tabular}}
	\centering\caption{\label{table3.average-energy} Minimum $\left[\left\langle E\right\rangle_{n,\ell,m}\right]_*$ and maximum $\left[\left\langle E\right\rangle_{n,\ell,m}\right]^*$ values of average energy (in eV) of molecules for $t_0=1973.29$ sec, $b=0.1$.}
\end{table}	

Here $\omega^{(f)}=\sqrt{\frac{2D_e}{\mu r_e^2}}$, and it follows that $\xi^{(c)}_{n,\ell}=\xi^{(f)}_{n+1,\ell}$. The average energy is a periodic function of $T$ with period $\sqrt{\frac{\pi^2\mu r_e^2}{2D_et_0^2}}$. One finds that, it is minimum if $T=2k\sqrt{\frac{\pi^2\mu r_e^2}{8D_et_0^2}}$ and maximum if $T=(2k+1)\sqrt{\frac{\pi^2\mu r_e^2}{8D_et_0^2}}$, where $k=0,1,2,\cdots$. Thus the average energy difference between two quantum states of $i$th molecule may expressed as follows, 
\beq
\ba{ll}
\left\langle\Delta  E\right\rangle^{(i,i)} (t)&=\left\langle E\right\rangle^{(i)}_{n_1,\ell_1,m_1}-\left\langle E\right\rangle^{(i)}_{n_2,\ell_2,m_2}\\
&=\ds \frac{\xi^{(c,i)}_{n_1,\ell_1}-\xi^{(c,i)}_{n_2,\ell_2}}{[\b^{(i)}]^2}+\ds\frac{2D^{(i)}_e}{[r^{(i)}_e]^2}\left(a+\frac{\mu^{(i)} [r^{(i)}_e]^2c_0^2}{2D^{(i)}_et_0^2[\b^{(i)}]^2}\right)\left[[r^{(i)}_{0,(n_1,\ell_1)}]^2\left\langle s^2\right\rangle^{(i)}_{n_1,\ell_1}-[r^{(i)}_{0,(n_2,\ell_2)}]^2\left\langle s^2\right\rangle^{(i)}_{n_2,\ell_2}\right],
\ea 
\eeq
and for two states $\{n_1, \ell_1, m_1\}$ and $\{n_2, \ell_2, m_2\}$ of molecules $i$ and $j$, it is given by, 
\beq
\ba{ll}
\left\langle\Delta  E\right\rangle^{(i,j)} (t)&=\left\langle E\right\rangle^{(i)}_{n_1,\ell_1,m_1}-\left\langle E\right\rangle^{(j)}_{n_2,\ell_2,m_2}\\
&=\ds \frac{\xi^{(c,i)}_{n,\ell}}{[\b^{(i)}]^2}+\ds\frac{\mu^{(i)} [r^{(i)}_{0,(n,\ell)}]^2}{2t_0^2}\left(\left[\dot{\b}^{(i)}\right]^2-\b^{(i)}\ddot{\b}^{(i)}\right)\left\langle s^2\right\rangle^{(i)}_{n,\ell}-\frac{\xi^{(c,j)}_{n,\ell}}{[\b^{(j)}]^2}-\frac{\mu [r^{(j)}_{0,(n,\ell)}]^2}{2t_0^2}\left(\left[\dot{\b}^{(j)}\right]^2-\b^{(j)}\ddot{\b}^{(j)}\right)\left\langle s^2\right\rangle^{(j)}_{n,\ell}.
\ea 
\eeq

\begin{figure}[h]
	\centering
	\includegraphics[width=19cm,height=14cm]{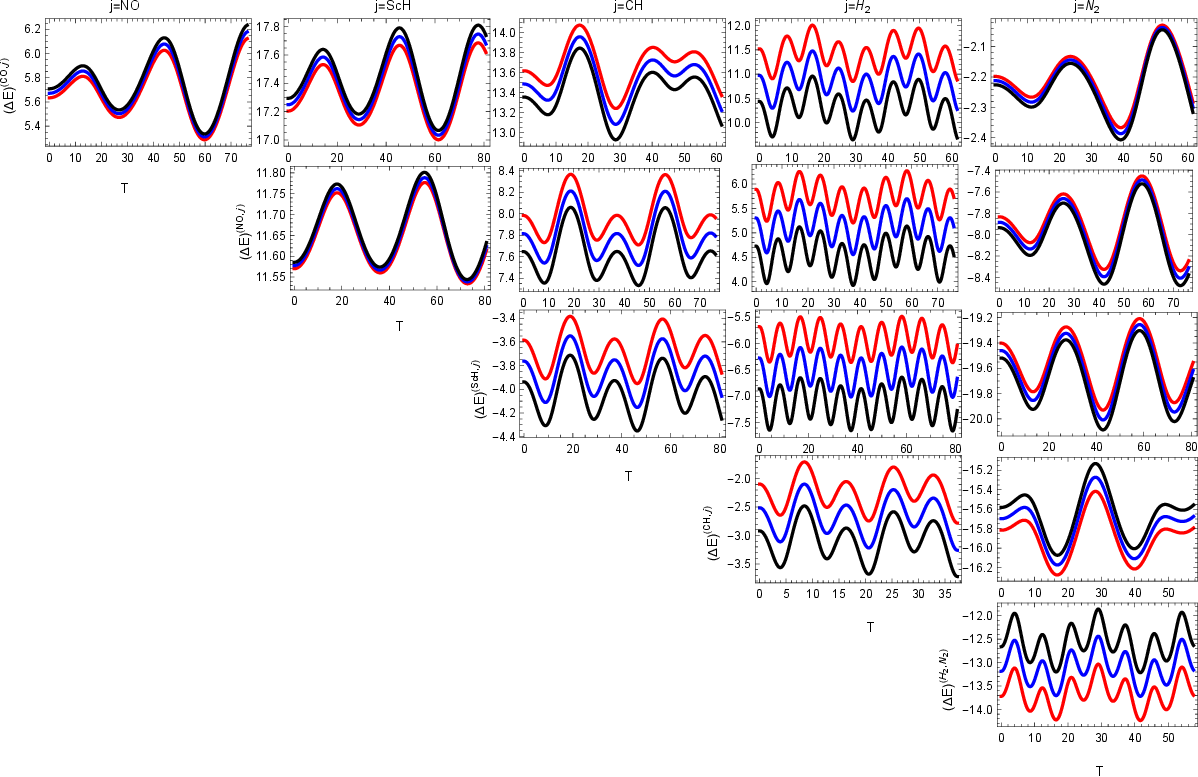}
	\caption{\label{fig3.average-energy-dif} Plot of average energy differences among a pair of molecules in eV units for $b=1$ and $t_0=$1973.29 sec. Red, blue and black lines are used for $n=1,\ell=0$; $n=2,\ell=1$; and $n=3,\ell=1$ states respectively.}
\end{figure}

The TD average energies of six molecules are plotted in Fig.~\ref{fig3.average-energy} for three states, namely, $n=1,\ell=0$; $n=2, \ell=1$ and $n=3, \ell=1$, with $b=0.1$ and $t_0=1000$ sec. This compares the average energy (solid lines) with ro-vibrational energy in the respective free quantum system (dashed line).  The average energies oscillate with time; moreover all the states in case of a given molecule show quite similar behaviour. At this point, it is interesting to examine the minimum and maximum values of the latter, which are reported in Table~\ref{table3.average-energy} for all the molecules, for six low-lying states. The respective average energy difference plots for all the molecules (with respect to other molecules) are given in Fig.~\ref{fig3.average-energy-dif}. Similarly, ro-vibrational energies with free boundary condition, for the molecules are presented in Table~\ref{table4.energy} for the same six states, along with the references in last column. 

The average energy at a particular time has been investigated in refs. \cite{rakhmanov2018,nath2020,nath2021,carbo-dorca2022,nath2022,nath2022a,nath2022_nld,nakamura2011,chen2010,mousavi2012}. Since the Hamiltonian is time dependent, the corresponding wave functions are necessarily time-dependent. We need average energy over a time domain. So, for comparison we will also investigate the average energy of the system over a time domain, instead of a fixed time. The former quantity over a time domain $[0,t]$ is given by \cite{pfeifer1993,luo2005,boykin2007,khalil2015}, 
	\beq
	\ba{ll}
	\widetilde{\left\langle E(t)\right\rangle}_{n,\ell,m}&=\ds\frac{1}{t}\int_0^t\left\langle E(t')\right\rangle_{n,\ell,m}dt'= \ds\frac{\mu r_{0,(n,\ell)}^2a \omega_1^2}{t_0^2} \left\langle s^2\right\rangle_{n,\ell} +\frac{ \xi_{n,\ell}^{(c)} +\frac{\mu r_{0,(n,\ell)}^2}{t_0^2} \omega_1^2 (b^2-a^2) \left\langle s^2\right\rangle_{n,\ell}}{\omega_1 \sqrt{a^2-b^2}}\left\{\frac{1}{t}\tan^{-1}\left[\sqrt{\frac{a-b}{a+b}}\tan(\omega_1t)\right]\right\}.
	\ea 
	\eeq
	We observe that $\lim\limits_{t\rightarrow0}\widetilde{\left\langle E(t)\right\rangle}_{n,\ell,m}=\ds \frac{\xi^{(c)}_{n,\ell}}{a+b}+\frac{2bD_er_{0,(n,\ell)}^2}{r_e^2}\left\langle s^2\right\rangle_{n,\ell}=\left\langle E(0)\right\rangle_{n,\ell,m}$.
	
	\begin{table}[t] 
		\centering
		\begin{tabular}{lllllllll}\hline\hline
			$n$&$\ell$ & \multicolumn{1}{c}{CO} & \multicolumn{1}{c}{NO}&  \multicolumn{1}{c}{ScH}& \multicolumn{1}{c}{CH}& \multicolumn{1}{c}{H$_2$}& \multicolumn{1}{c}{N$_2$}&Ref. \\\hline
			1&0&21.9959&16.3349&4.73348&8.39998&10.6261&24.2038&\cite{oyewumi2012}\\
			2&0&22.1997&16.4998&4.88902&8.73644&11.3827&24.4221&\\
			2&1&22.2001&16.5002&4.89037&8.74003&11.3978&24.4226&\\
			3&0&22.4034&16.6647&5.04456&9.0729&12.1394&24.6403&\\
			3&1&22.4039&16.6651&5.04591&9.07649&12.1545&24.6408&\\
			3&2&22.4049&16.666&5.04859&9.08365&12.1845&24.6418&\\\hline\hline
		\end{tabular}
		\centering\caption{\label{table4.energy} Ro-vibrational energy $\xi^{(c)}_{n,\ell}$ of diatomic molecules (in eV) in a confined system in presence of pseudoharmonic oscillator and its comparison with the respective free system, $\ds\xi^{(f)}_{n+1,\ell}-2D_e$.}
	\end{table}
		
\begin{figure}[t]
	\centering
	\includegraphics[width=19cm,height=12cm]{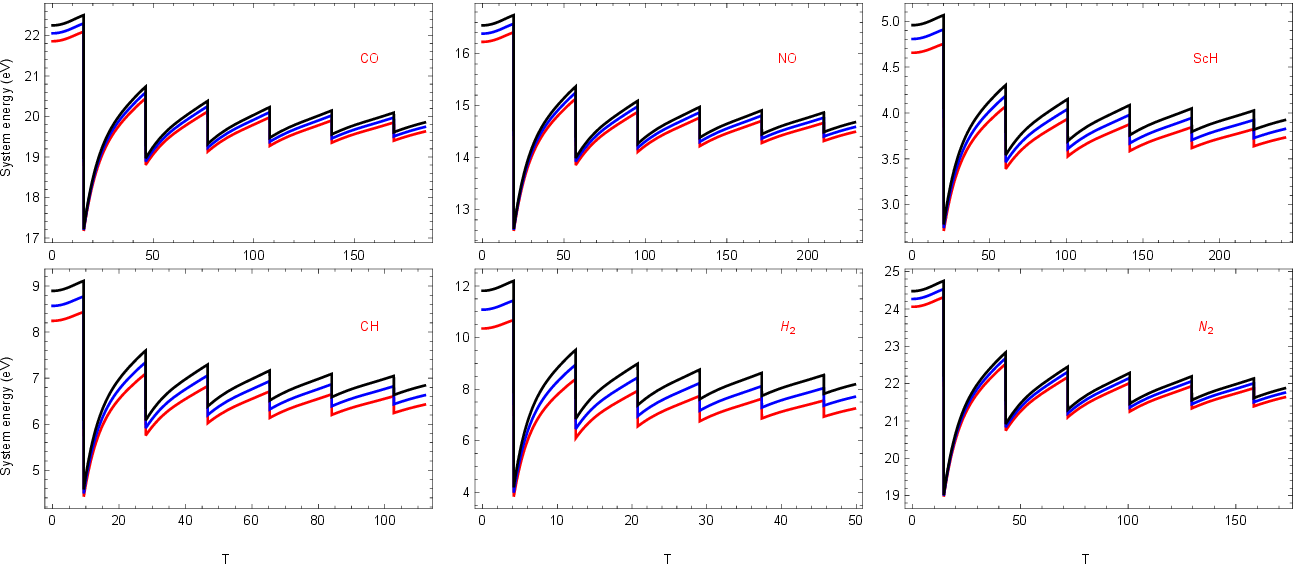}
	\caption{\label{fig3.system-energy} Plot of the TD system energy (in eV) of molecules, for $b=1$ and $t_0=1973.29\,$sec. Solid and dashed lines represent $\xi^{(c)}_{n,\ell}$ and average energy respectively. Red, blue and black lines are drawn for $n=1,\ell=0$; $n=2,\ell=1$; and $n=3,\ell=1$ states respectively.}
\end{figure}

The system energy decays from the initial state. It is piecewise monotone increasing function of time but not periodic. It has infinitely many local minimum and maximum values. We observe that at time  $t^*_{k}=\frac{(2k-1)\pi}{2\om_1},~k=1,2,3,\cdots$, the energy $\widetilde{\left\langle E\right\rangle}_{n,\ell,m}(t)$ is discontinuous, it is left continuous and at this point it is maximum such that,   
\beq
\lim\limits_{t\rightarrow t^{*}_k-0}\widetilde{\left\langle E(t)\right\rangle}_{n,\ell,m}=\widetilde{\left\langle E(t^*_k)\right\rangle}_{n,\ell,m}\ne\lim\limits_{t\rightarrow t^*_k+0}\widetilde{\left\langle E(t)\right\rangle}_{n,\ell,m}.
\eeq 
The sequences of local maximum and minimum values of the energy $\widetilde{\left\langle E\right\rangle}_{n,\ell,m}(t)$ are obtained as, 
\beq
\ba{ll}
\left[\widetilde{\left\langle E\right\rangle}_{n,\ell,m}\right]_k^*=\ds\frac{\mu r_{0,(n,\ell)}^2a \omega_1^2}{t_0^2} \left\langle s^2\right\rangle_{n,\ell}+\frac{ \xi_{n,\ell}^{(c)} +\frac{\mu r_{0,(n,\ell)}^2}{t_0^2} \omega_1^2 (b^2-a^2) \left\langle s^2\right\rangle_{n,\ell}}{(2k-1)\sqrt{a^2-b^2}},~~\left[\widetilde{\left\langle E\right\rangle}_{n,\ell,m}\right]_k^*<\left[\widetilde{\left\langle E\right\rangle}_{n,\ell,m}\right]_{k+1}^*,~k=1,2,3,\cdots,\\
\left[\widetilde{\left\langle E\right\rangle}_{n,\ell,m}\right]_{k*}=\ds\frac{\mu r_{0,(n,\ell)}^2a \omega_1^2}{t_0^2} \left\langle s^2\right\rangle_{n,\ell}-\frac{ \xi_{n,\ell}^{(c)} +\frac{\mu r_{0,(n,\ell)}^2}{t_0^2} \omega_1^2 (b^2-a^2) \left\langle s^2\right\rangle_{n,\ell}}{(2k-1)\sqrt{a^2-b^2}},~~\left[\widetilde{\left\langle E\right\rangle}_{n,\ell,m}\right]_{k*}>\left[\widetilde{\left\langle E\right\rangle}_{n,\ell,m}\right]_{k+1*},~k=1,2,3,\cdots.
\ea
\eeq
The sequence of local maximum values $\left\{\left[\widetilde{\left\langle E\right\rangle}_{n,\ell,m}\right]_k^*\right\}_{k=1}^{\infty}$ is monotone decreasing and the sequence of local minimum values $\left\{\left[\widetilde{\left\langle E\right\rangle}_{n,\ell,m}\right]_{k*}\right\}_{k=1}^{\infty}$ is monotone increasing. Both these two sequences converge to the same point, i.e.,
\beq
\lim\limits_{k\rightarrow\infty}\left[\widetilde{\left\langle E\right\rangle}_{n,\ell,m}\right]_k^*=\lim\limits_{k\rightarrow\infty}\left[\widetilde{\left\langle E\right\rangle}_{n,\ell,m}\right]_{k*}=\ds\frac{2ar_{0,(n,\ell)}^2 D_e}{r_e^2} \left\langle s^2\right\rangle_{n,\ell}.
\eeq
Therefore, we can say that after a long time, the energy $\widetilde{\left\langle E(t)\right\rangle}_{n,\ell,m}$ collapses to the value $\ds\frac{2ar_{0,(n,\ell)}^2 D_e}{r_e^2} \left\langle s^2\right\rangle_{n,\ell}$. In Fig. \ref{fig3.system-energy} we have plotted the average energy $\widetilde{\left\langle E(t)\right\rangle}_{n,\ell,m}$ over a time domain $[0,t]$. From this figure it is clear that $\widetilde{\left\langle E(t)\right\rangle}_{n,\ell,m}$ is oscillating but not periodic and the length of oscillation decreases as time increases. After a sufficiently long time, it attains a finite value.      

Since, the non-interacting particles are confined in a spherical box of TD radius, the boundary wall moves with some velocity. We note that, the force acting on a fixed boundary in a time-independent Hamiltonian is given by $F=-\frac{\partial E}{\partial L}$, where $L$ is the fixed length of the cavity and $E$ is the energy of a state $\psi$. But for a TD Hamiltonian with moving boundary wall having a TD length $L(t)$, the average force is defined by $\left\langle F\right\rangle=-\frac{\partial \left\langle E\right\rangle}{\partial L(t)}$.
 One can find the TD average force acting on the wall, as given by: $\left\langle F(t)\right\rangle_{n,\ell,m}=-\frac{\partial \left\langle E(t)\right\rangle}{\partial r_{0,(n,\ell)} \beta}$. It can be expressed in closed analytical form as \cite{nakamura2011,chen2010,mousavi2012},  
	\beq
	\left\langle F(t)\right\rangle_{n,\ell,m}=\ds \frac{2 \xi^{(c)}_{n,\ell}}{r_{0,(n,\ell)} \b^3}+\ds\frac{\mu r_{0,(n,\ell)}}{2t_0^2}\ddot{\b}\left\langle s^2\right\rangle_{n,\ell},
	\eeq 
	The average force is a periodic and oscillating function of time. It has maximum and minimum values. The maximum average force is found to be, 
	\beq
	[\left\langle F\right\rangle_{n,\ell,m}]^*=\left\{  \frac{2 \xi^{(c)}_{n,\ell}}{r_{0,(n,\ell)} }+\ds\frac{\mu r_{0,(n,\ell)} c_0^2}{2t_0^2}\left\langle s^2\right\rangle_{n,\ell}\right\} \frac{1}{(a-b)^{3/2}}-\frac{ D_e r_{0,(n,\ell)}}{r_e^2}\left\langle s^2\right\rangle_{n,\ell}\sqrt{a-b},
	\eeq 
	and the minimum value is obtained as, 
	\beq
	[\left\langle F\right\rangle_{n,\ell,m}]_*=\left\{  \frac{2 \xi^{(c)}_{n,\ell}}{r_{0,(n,\ell)} }+\ds\frac{\mu r_{0,(n,\ell)} c_0^2}{2t_0^2}\left\langle s^2\right\rangle_{n,\ell}\right\} \frac{1}{(a+b)^{3/2}}-\frac{ D_e r_{0,(n,\ell)}}{r_e^2}\left\langle s^2\right\rangle_{n,\ell}\sqrt{a+b}.
	\eeq
	Table \ref{table3.average-force} reports the minimum and maximum values of average force acting on the boundary wall for six molecules.
	
		 Next, the corresponding average pressure obtained by the moving boundary wall is given by $\left\langle P\right\rangle=\frac{\left\langle F\right\rangle}{4\pi L^2(t)}$. The time-dependence of the boundary wall is considered as periodically breathing radius, where the TD scale factor of the internuclear distance is obtained from the Ermakov equation.

The average pressure, $\left\langle P(t)\right\rangle_{n,\ell,m}=\frac{\left\langle F(t)\right\rangle}{4 \pi r_{0,(n,\ell)}^2 \beta^2}$, can be expressed in analytical form as, 
	\beq
	\left\langle P(t)\right\rangle_{n,\ell,m}=\ds \frac{1}{4 \pi r_{0,(n,\ell)}^2}\left[\ds \frac{2 \xi^{(c)}_{n,\ell}}{r_{0,(n,\ell)} \b^5}+\ds\frac{\mu r_{0,(n,\ell)}}{2t_0^2}\frac{\ddot{\b}}{\b^2}\left\langle s^2\right\rangle_{n,\ell} \right].
	\eeq
	The minimum and maximum values of the pressure are obtained as, 
	\beq
	\ba{l}
	\left[\left\langle P\right\rangle_{n,\ell,m}\right]^*=\ds \frac{1}{4 \pi r_{0,(n,\ell)}^2}\left[\left\{  \frac{2 \xi^{(c)}_{n,\ell}}{r_{0,(n,\ell)} }+\ds\frac{\mu r_{0,(n,\ell)} c_0^2}{2t_0^2}\left\langle s^2\right\rangle_{n,\ell}\right\} \frac{1}{(a-b)^{5/2}}-\frac{ D_e r_{0,(n,\ell)}}{r_e^2\sqrt{a-b}}\left\langle s^2\right\rangle_{n,\ell} \right],\\
	\left[\left\langle P\right\rangle_{n,\ell,m}\right]_*=\ds \frac{1}{4 \pi r_{0,(n,\ell)}^2}\left[\left\{  \frac{2 \xi^{(c)}_{n,\ell}}{r_{0,(n,\ell)} }+\ds\frac{\mu r_{0,(n,\ell)} c_0^2}{2t_0^2}\left\langle s^2\right\rangle_{n,\ell}\right\} \frac{1}{(a+b)^{5/2}}-\frac{ D_e r_{0,(n,\ell)}}{r_e^2\sqrt{a+b}}\left\langle s^2\right\rangle_{n,\ell} \right].
	\ea
	\eeq
	In Table \ref{table3.average-pressure}, we have provided the minimum and maximum values of the pressure. 
			\begin{table}[h]
			\centering
			\scalebox{.9}{\begin{tabular}{llllllllllllll}\hline\hline
					$n$&$\ell$ & \multicolumn{2}{c}{CO} & \multicolumn{2}{c}{NO}&  \multicolumn{2}{c}{ScH}& \multicolumn{2}{c}{CH}& \multicolumn{2}{c}{H$_2$}& \multicolumn{2}{c}{N$_2$}\\ \hline
					&&\multicolumn{1}{c}{Min}&\multicolumn{1}{c}{Max} &\multicolumn{1}{c}{Min}&\multicolumn{1}{c}{Max}&\multicolumn{1}{c}{Min}&\multicolumn{1}{c}{Max}&\multicolumn{1}{c}{Min}&\multicolumn{1}{c}{Max}&\multicolumn{1}{c}{Min}&\multicolumn{1}{c}{Max}&\multicolumn{1}{c}{Min}&\multicolumn{1}{c}{Max}\\\hline
					1&0&31.8455&46.9971&23.1891&34.2044&4.35278&6.35651&12.2431&17.8366 & 23.2864&33.6394&36.1399&53.343\\
					
					2&0&31.0304&45.78&22.5778&33.2919&4.19665&6.12486&11.7924&17.1693 & 22.427&32.3811&35.2235&51.9746\\
					
					2&1&31.0307&45.7805&22.5781&33.2923&4.19723&6.12572&11.7948&17.1729 & 22.4413&32.4024&35.2238&51.9752\\
					
					3&0&30.5255&45.021&22.2026&32.7279&4.12048&6.01073&11.5887&16.8645 & 22.1733&32.0069&34.6544&51.1189\\
					
					3&1&30.5258&45.0215&22.2029&32.7284&4.12104&6.01157&11.591&16.868 & 22.1869&32.0272&34.6547&51.1195\\
					
					3&2&30.5266&45.0224&22.2035&32.7292&4.12216&6.01324&11.5956&16.8749 & 22.2139&32.0676&34.6554&51.1206\\\hline
					\hline
			\end{tabular}}
			\centering\caption{\label{table3.average-force} Minimum $\left[\left\langle F\right\rangle_{n,\ell,m}\right]_*$ and maximum $\left[\left\langle F\right\rangle_{n,\ell,m}\right]^*$ values of average force (in eV/sec) of molecules for $t_0=1973.29$ sec, $b=0.1$.}
		\end{table}
		
	\section{Time-correlation function}\label{time-co}
\begin{figure}[]
	\centering
	\includegraphics[width=18cm,height=11cm]{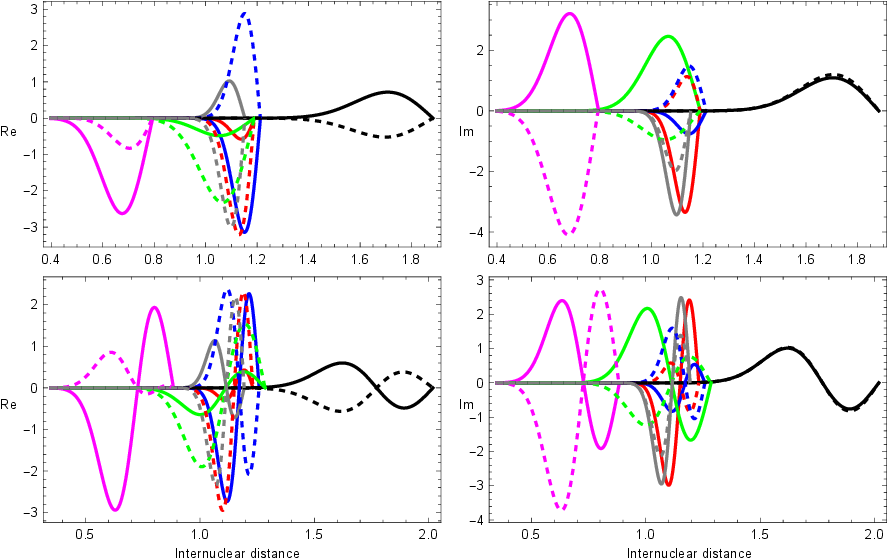}
	\caption{\label{fig.sol} Comparison of real and imaginary parts of radial wave functions for two different time domains of six diatomic molecules, for $b=0.1$ and $t_0= $1973.29 sec. The red, blue, black, green, magenta and gray represent plots for CO, NO, ScH, CH, H$_2$, N$_2$. Solid and dashed lines refer to $t=0.5$ and $t=1.0$ respectively. The top and bottom panels refer to ground $(n=1)$ with $\ell=0$ and first excited state ($n=2$) with $\ell=1$. }
\end{figure}

Now, let us consider the radial wave function, which is complex in the TD confined quantum system. The real and imaginary parts are plotted in Fig.~\ref{fig.sol} for two time domains, for all the 6 molecules. The solid and dashed curves signify $t=0.5$ and $t=1$ respectively. The upper and lower panels correspond to ground state ($n=1$) with $\ell =0$ and first excited state ($n=2$) with $\ell=1$. These figures also illustrate the impenetrability of the well. 

	\begin{table}[]
	\centering
	\scalebox{.9}{\begin{tabular}{llllllllllllll}\hline\hline
			$n$&$\ell$ & \multicolumn{2}{c}{CO} & \multicolumn{2}{c}{NO}&  \multicolumn{2}{c}{ScH}& \multicolumn{2}{c}{CH}& \multicolumn{2}{c}{H$_2$}& \multicolumn{2}{c}{N$_2$}\\ \hline
			&&\multicolumn{1}{c}{Min}&\multicolumn{1}{c}{Max} &\multicolumn{1}{c}{Min}&\multicolumn{1}{c}{Max}&\multicolumn{1}{c}{Min}&\multicolumn{1}{c}{Max}&\multicolumn{1}{c}{Min}&\multicolumn{1}{c}{Max}&\multicolumn{1}{c}{Min}&\multicolumn{1}{c}{Max}&\multicolumn{1}{c}{Min}&\multicolumn{1}{c}{Max}\\\hline
			1&0&1.79306&3.23097&1.25457&2.25948&0.0976913&0.17419&0.688433&1.22461 & 2.9318&5.17124&2.16473&3.9013\\
			
			2&0&1.62798&2.93259&1.13449&2.04255&0.082014&0.146149&0.568297&1.01028 & 2.28107&4.02137&1.96785&3.54541\\
			
			2&1&1.62796&2.93256&1.13448&2.04253&0.0820029&0.14613&0.56818&1.01008 & 2.27944&4.01858&1.96783&3.54537\\
			
			3&0&1.52119&2.73937&1.05723&1.90282&0.0728746&0.129799&0.499811&0.888099 & 1.93773&3.41523&1.84029&3.31456\\
			
			3&1&1.52117&2.73934&1.05721&1.9028&0.0728658&0.129783&0.49972&0.887942 & 1.93653&3.41319&1.84027&3.31453\\
			
			3&2&1.52115&2.73928&1.05719&1.90276&0.0728482&0.129753&0.499538&0.887629 & 1.93414&3.40913&1.84024&3.31447\\\hline
			\hline
	\end{tabular}}
	\centering\caption{\label{table3.average-pressure} Minimum $\left[\left\langle P\right\rangle_{n,\ell,m}\right]_*$ and maximum $\left[\left\langle P\right\rangle_{n,\ell,m}\right]^*$ values of average pressure (in eV) of molecules for $t_0=1973.29$ sec, $b=0.1$.}
\end{table}

For two states denoted by $\psi^{(i)}_{n_1,\ell_1,m_1}$ and $\psi^{(j)}_{n_2,\ell_2,m_2}$, the time correlation function can be expressed as \cite{tokmakoff2014,fring2020,nath2022}, 
\beq
\ba{ll}\label{corr}
\mathcal{C}^{(i,j)}_{(n_1,\ell_1,m_1),(n_2,\ell_2,m_2)}(t_1,t_2)&=\ds\int\left[\psi^{(i)}_{n_1,\ell_1,m_1}({\bf r},t_1)\right]^*\psi^{(j)}_{n_2,\ell_2,m_2}({\bf r},t_2)d{\bf r}\\
&=\ds\int_0^{\widetilde{\b}^{(i,j)}_{12}} \left[R^{(i)}_{n_1,\ell_1}(r,t_1)\right]^*R^{(j)}_{n_2,\ell_2}(r,t_2)\,r^2dr\int_{\theta=0}^{\pi}\int_{\phi=0}^{2\pi}Y^*_{\ell_1,m_1}(\theta,\phi)Y_{\ell_2,m_2}(\theta,\phi)\sin\theta d\theta d\phi,
\ea 
\eeq 
where $\left(0,\widetilde{\b}^{(i,j)}_{12}\right)$ is the common domain of the overlap integral of moving boundary solutions, and $\widetilde{\b}^{(i,j)}_{12}=\min\left\{r^{(i)}_{0,(n_1,\ell_1)}\b^{(i)}(T_1),\right.$ $\left.\,r^{(j)}_{0,(n_2,\ell_2)}\b^{(j)}(T_2)\right\}$. In our problem, we focus on the radial density functions which are dependent on time. Therefore time correlation function can be obtained from the corresponding radial wave functions $R^{(i)}_{n_1,\ell}(r,t_1)$ and $R^{(j)}_{n_2,\ell}(r,t_2)$. 

\subsection{Cross-correlation function}
The quantity $\mathcal{C}^{(i,j)}_{(n_1,\ell,m),(n_2,\ell,m)}(t_1,t_2)=\mathcal{C}^{(i,j)}_{(n_1,n_2,\ell,m)}(t_1,t_2)$ is called the cross-correlation function connecting the two states $\psi^{(i)}_{n_1,\ell,m}({\bf r},t_1)$ and $\psi^{(j)}_{n_2,\ell,m}({\bf r},t_2)$ at two different times $t_1$ and $t_2$. For central potentials, this function, $\mathcal{C}^{(i,j)}_{(n_1,n_2,\ell,m)}(t_1,t_2)$, symbolized as $C^{(i,j)}_{(n_1,n_2,\ell)}(t_1,t_2)$, is independent of $\ell$ and as such, is given by, 
\beq
\ba{ll}
\mathcal{C}^{(i,j)}_{(n_1,n_2,\ell)}(t_1,t_2)&=\ds\int_0^{\widetilde{\b}^{(i,j)}_{12}} \left[R^{(i)}_{n_1,\ell}(r,t_1)\right]^*R^{(j)}_{n_2,\ell}(r,t_2)\,r^2dr.
\ea 
\eeq
For the current problem, the cross-correlation function, $\mathcal{C}^{(i,j)}_{(n_1,n_2,\ell)}(t_1,t_2)$ is a function of $IDR^{(i,j)}_{n_1,n_2,\ell}(T_1,T_2)$, at two different time domains. 
Using this and after some algebraic manipulation, one obtains the analytical form of $C^{(i,j)}_{(n_1,n_2,\ell)}(t,0)$ as, 
\beq
\ba{ll}
\mathcal{C}^{(i,j)}_{(n_1,n_2,\ell)}(t,0)
&=\ds\left(\frac{[r_{0,(n_1,\ell)}^{(i)}r_{0,(n_2,\ell)}^{(j)}]^{\frac{3}{2}}N_{n_1,\ell}^{(i)}N_{n_2,\ell}^{(j)}[IDR^{(i,j)}_{n_1,n_2,\ell}(T,0)]^{L^{(j)}+\frac{3}{2}}\,2^{\frac{L^{(i)}+L^{(j)}+1}{2}}\exp\left\{if^{(c,i)}_{n_1,\ell}(T)\right\}}{\left[\lambda_{n_1,\ell}^{(i)}+\widetilde{\lambda}_{n_1,n_2,\ell}^{(i,j)}(T,0)\right]^{\frac{L^{(i)}+L^{(j)}+3}{2}}}\right)\\\\
&\ds\quad\times\left(\sum\limits_{k=0}^{n_1}\sum\limits_{l=0}^{n_2}\left\{\left[\frac{2\lambda_{n_1,\ell}^{(i)}}{\lambda_{n_1,\ell}^{(i)}+\widetilde{\lambda}_{n_1,n_2,\ell}^{(i,j)}(T,0)}\right]^k\left[\frac{2\widetilde{\lambda}_{n_1,n_2,\ell}^{(i,j)}}{\lambda_{n_1,\ell}^{(i)}+\widetilde{\lambda}_{n_1,n_2,\ell}^{(i,j)}(T,0)}\right]^l\left[\frac{(-n_1)_k(-n_2)_l}{(L^{(i)}+\frac{3}{2})_k\,k!(L^{(j)}+\frac{3}{2})_l\,l!}\right]\right.\right.\\\\
&\quad \left.\left.\times \ds\left[\mathcal{J}_c\left(\ba{cc}\frac{L^{(i)}+L^{(j)}+1}{2}+k+l,&\frac{\alpha^{(i)}_{n_1,\ell}(T)}{\lambda_{n_1,\ell}^{(i)}+\widetilde{\lambda}_{n_1,n_2,\ell}^{(i,j)}(T,0)}\\0,&x^{(i,j)}_{n_1,n_2,\ell}(T,0)\ea\right)+i\,\mathcal{J}_s\left(\ba{cc}\frac{L^{(i)}+L^{(j)}+1}{2}+k+l,&\frac{\alpha^{(i)}_{n_1,\ell}(T)}{\lambda_{n_1,\ell}^{(i)}+\widetilde{\lambda}_{n_1,n_2,\ell}^{(i,j)}(T,0)}\\
0,&x^{(i,j)}_{n_1,n_2,\ell}(T,0)\ea\right)\right]\right\}\right),
\ea
\eeq
where
\beq
\ba{ll}
\ds \ds f^{(c,i)}_{n_1,\ell}(T)=\ds\frac{t_0\xi^{(c,i)}_{n_1,\ell}}{\sqrt{a^2-b^2}\hbar\om^{(i)}}\tan^{-1}\left(\sqrt{\frac{a-b}{a+b}}\tan(\om^{(i)}T)\right),~
\om^{(i)}=\ds\sqrt{\frac{2D_e^{(i)}t_0^2}{\mu^{(i)} [r_e^{(i)}]^2}},~
\ds\alpha^{(i)}_{n_1,\ell}=\ds\frac{\mu^{(i)}[r^{(i)}_{0,(n_1,\ell)}]^2\beta^{(i)}(T)\dot{\beta}^{(i)}(T)}{\hbar t_0},\\
\ds\widetilde{\lambda}_{n_1,n_2,\ell}^{(i,j)}(T_1,T_2)=\lambda_{n_2,\ell}^{(j)}\left[IDR^{(i,j)}_{n_1,n_2,\ell}(T_1,T_2)\right]^2,~
x^{(i,j)}_{n_1,n_2,\ell}(T_1,T_2)=\ds\min\left\{\frac{\lambda_{n_1,\ell}^{(i)}+\widetilde{\lambda}_{n_1,n_2,\ell}^{(i,j)}(T_1,T_2)}{2},\frac{\lambda_{n_1,\ell}^{(i)}+\widetilde{\lambda}_{n_1,n_2,\ell}^{(i,j)}(T_1,T_2)}{2\left[IDR^{(i,j)}_{n_1,n_2,\ell}(T_1,T_2)\right]^2}\right\},\\
\ds L^{(i)}(\ell)=-\frac{1}{2}+\sqrt{\left(\ell+\frac{1}{2}\right)^2+\frac{2\mu^{(i)} D_e^{(i)}[r_e^{(i)}]^2}{\hbar^2}},

a^{(i)}=\sqrt{b^2+\frac{\mu^{(i)} [r_e^{(i)}]^2c_0^2}{2D_e^{(i)} t_0^2}}.
\ea
\eeq
and
\beq
\ba{ll}
\ds\mathcal{J}_c\left(\ba{cc}\gamma,&\d\\\ze,&\eta\ea\right)=\ds\int_\zeta^\eta e^{-z}z^\gamma\cos(\delta z)dz,~
\ds\mathcal{J}_s\left(\ba{cc}\gamma,&\delta\\\zeta,&\eta\ea\right)=\ds\int_\zeta^\eta e^{-z}z^\gamma\sin(\delta z)dz.
\ea
\eeq 

\begin{figure}[t]
		\centering
		\includegraphics[width=19cm,height=14cm]{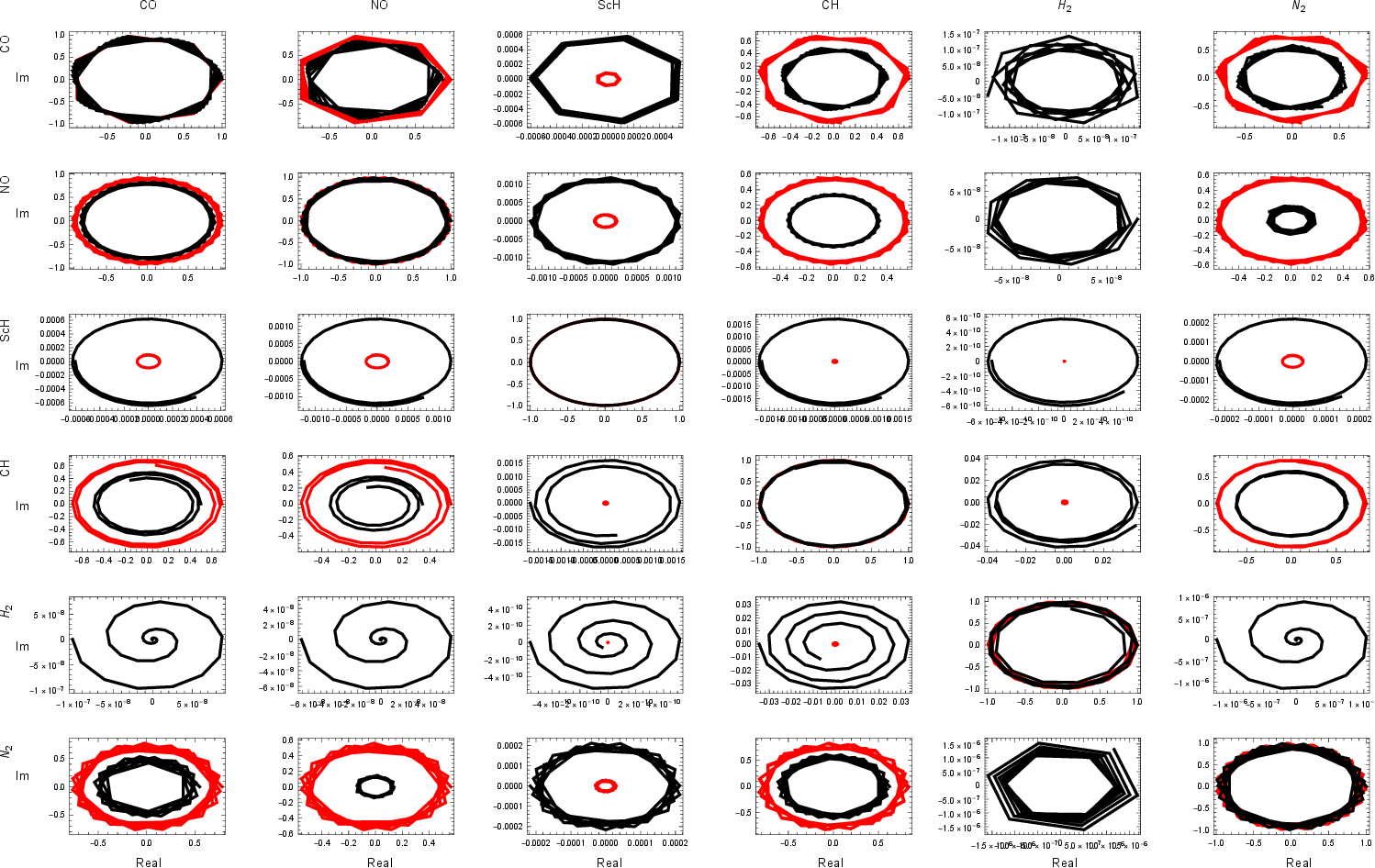}
		\caption{\label{fig7.cross-correlation} Plot of time correlation function among pairs of diatomic molecules, for $b=0.1$, $t_0=1973.29$ sec. The red and black curves represent $n=1,\ell=0$ and $n=2,\ell=1$ states respectively.}
	\end{figure}

It can be seen that, the time-correlation function between two molecules at different time domains $t_1$, $t_2$ is a function of the corresponding $IDR^{(i,j)}_{n_1,n_2,\ell}(T_1,T_2)$. We have calculated the numerical values of $\mathcal{C}^{(i,j)}_{(n_1,n_2,\ell)}(t,0)$ for some pairs of molecules for representative $b=0.1$ and $t_0=1973.29$ sec. These are depicted in Fig.~\ref{fig7.cross-correlation} for all pairs arising out of the the six molecules, for $n=1, \ell=0$ and $n=2, \ell=1$ states respectively, which are marked in red and black colors. It may be noted that, time correlation function and average energy of diatomic molecules (H$_2$, LiH, HCl, ScH, TiH, CrH, VH, CO) in Deng-Fan potential in a moving boundary has been studied in \cite{nath2022_nld}.

The simplest form of $\mathcal{C}^{(i,j)}_{(n_1,n_2,\ell)}(t,0)$ is a function of $IDR^{(i,j)}_{n_1,n_2,\ell}(T,0)$. For $0<\gamma<1$, we have used numerical methods to compute the values of integrals $\mathcal{J}_c\left(\ba{cc}\gamma,&\delta\\\zeta,&\eta\ea\right)$ and $\mathcal{J}_s\left(\ba{cc}\gamma,&\delta\\\zeta,&\eta\ea\right)$. For $\gamma>1$, one can use the following combined recursion relations, 
\beq
\left.\ba{ll}
\ds\mathcal{J}_c\left(\ba{cc}\gamma,&\delta\\\zeta,&\eta\ea\right)=\ds\left[\frac{e^{-z}z^\gamma}{\delta^2+1}\left(\delta\sin(\delta z)-\cos(\delta z)\right)\right]_\zeta^\eta+\frac{\gamma}{\delta^2+1}\mathcal{J}_c\left(\ba{cc}\gamma-1,&\delta\\\zeta,&\eta\ea\right)-\frac{\gamma\delta}{\delta^2+1}\mathcal{J}_s\left(\ba{cc}\gamma-1,&\delta\\\zeta,&\eta\ea\right),\\\\
\ds\mathcal{J}_s\left(\ba{cc}\gamma,&\delta\\\zeta,&\eta\ea\right)=\ds\left[-\frac{e^{-z}z^\gamma}{\delta^2+1}\left(\sin(\delta z)+\delta\cos(\delta z)\right)\right]_\zeta^\eta+\frac{\gamma}{\delta^2+1}\mathcal{J}_s\left(\ba{cc}\gamma-1,&\delta\\\zeta,&\eta\ea\right)+\frac{\gamma\delta}{\delta^2+1}\mathcal{J}_c\left(\ba{cc}\gamma-1,&\delta\\\zeta,&\eta\ea\right),
\ea \right\}.
\eeq 
If $~\gamma\in\mathbb{N}$, then these two integrals can be performed analytically. For the current problem, $\frac{L^{(i)}+L^{(j)}+1}{2}+k+l,$ is always a non-integral positive real number. When $\frac{L^{(i)}+L^{(j)}+1}{2}+k+l>1$, we can use the above recursion formula and then apply numerical techniques for $0<\frac{L^{(i)}+L^{(j)}+1}{2}+k+l<1$. 

If $b=0$, then the cross-correlation function $\ds \mathcal{C}^{(i,j)}_{(n_1,n_2,\ell)}(t_1,t_2)$ reduces to $\ds\exp\left\{\frac{i\xi_{n_1,\ell}^{(c,i)}t_1}{\hbar a^{(i)}}-\frac{i\xi_{n_2,\ell}^{(c,j)}t_2}{\hbar a^{(j)}}\right\}$ times a simple overlap integral between two molecular states.
	
\subsection{Auto-correlation function}\label{autocorrelation}
The quantity $\mathcal{C}^{(i,i)}_{(n,n,\ell,m)}(t_1,t_2)$ is called auto-correlation function of $\psi^{(i)}_{n,\ell,m}$ at two different times $t=t_1$ and $t=t_2$. This is also independent of $m$, and denoted as $\mathcal{C}^{(i)}_{(n,\ell)}(t_1,t_2)$. Here we have been able to calculate auto-correlation $\mathcal{C}^{(i)}_{(n,\ell)}(t,0)$ analytically, which is obtained from the above cross-relation, and derived as, 
\beq
\ba{ll}
\mathcal{C}^{(i)}_{(n,\ell)}(t,0)&=\ds\frac{[r_{0,(n,\ell)}^{(i)}]^3[N_{n,\ell}^{(i)}]^2[IDR^{(i,i)}_{n,n,\ell}(T,0)]^{L^{(i)}+\frac{3}{2}}\,2^{L^{(i)}+\frac{1}{2}}\exp\left\{if^{(c,i)}_{n,\ell}(T)\right\}}{\left[\lambda_{n,\ell}^{(i)}+\widetilde{\lambda}_{n,n,\ell}^{(i,i)}(T,0)\right]^{L^{(i)}+\frac{3}{2}}}\\\\
&\quad\times\ds\sum\limits_{k=0}^{n}\sum\limits_{l=0}^{n}\left\{\left[\frac{2\lambda_{n,\ell}^{(i)}}{\lambda_{n,\ell}^{(i)}+\widetilde{\lambda}_{n,n,\ell}^{(i,i)}(T,0)}\right]^k\left[\frac{2\widetilde{\lambda}_{n,n,\ell}^{(i,i)}(T,0)}{\lambda_{n,\ell}^{(i)}+\widetilde{\lambda}_{n,n,\ell}^{(i,i)}(T,0)}\right]^l\frac{(-n)_k(-n)_l}{(L^{(i)}+\frac{3}{2})_k(L^{(i)}+\frac{3}{2})_l}\right.\\\\
&\quad \left.\times \ds\left[\mathcal{J}_c\left(\ba{cc}L^{(i)}+\frac{1}{2}+k+l,&\frac{\alpha^{(i)}_{n,\ell}(T)}{\lambda_{n,\ell}^{(i)}+\widetilde{\lambda}_{n,n,\ell}^{(i,i)}(T,0)}\\0,&x^{(i,i)}_{n,n,\ell}(T,0)\ea\right)+i\,\mathcal{J}_s\left(\ba{cc}L^{(i)}+\frac{1}{2}+k+l,&\frac{\alpha^{(i)}_{n,\ell}(T)}{\lambda_{n,\ell}^{(i)}+\widetilde{\lambda}_{n,n,\ell}^{(i,i)}(T,0)},\\
0,&x^{(i,i)}_{n,n,\ell}(T,0)\ea\right)\right]\right\}.
\ea 	
\eeq 
The auto-correlation function satisfies the relation $c_{n,l}^{i} (0,t)=[c_{n,l}^{i} (t,0)]^{*}$.\\
If $b=0$, then  $\mathcal{C}^{(i)}_{(n,\ell)}(t_1,t_2)$ reduces to $\ds\exp\left\{\frac{i\xi_{n_1,\ell}^{(c,i)}t_1}{\hbar a^{(i)}}-\frac{i\xi_{n_2,\ell}^{(c,j)}t_2}{\hbar a^{(j)}}\right\}$. The figures in principal diagonal of  Fig. \ref{fig7.cross-correlation} represent the auto-correlation functions of CO, NO, ScH, CH, H$_2$ and N$_2$ molecules respectively. 

\begin{table}[t]
	\centering
	\scalebox{.9}{\begin{tabular}{ll|ll|ll|ll|ll|ll|ll}\hline\hline
			$n$&$\ell$ & \multicolumn{2}{c|}{CO} & \multicolumn{2}{c|}{NO}&  \multicolumn{2}{c|}{ScH}& \multicolumn{2}{c|}{CH}& \multicolumn{2}{c|}{H$_2$}& \multicolumn{2}{c}{N$_2$}\\ \hline
			&
			&\multicolumn{1}{c}{$\mathcal{P}_{n,\ell}(t)$}&\multicolumn{1}{c|}{$[\mathcal{P}_{n,\ell}(t)]_*$} &\multicolumn{1}{c}{$\mathcal{P}_{n,\ell}(t)$}&\multicolumn{1}{c|}{$[\mathcal{P}_{n,\ell}(t)]_*$}
			&\multicolumn{1}{c}{$\mathcal{P}_{n,\ell}(t)$}&\multicolumn{1}{c|}{$[\mathcal{P}_{n,\ell}(t)]_*$}
			&\multicolumn{1}{c}{$\mathcal{P}_{n,\ell}(t)$}&\multicolumn{1}{c|}{$[\mathcal{P}_{n,\ell}(t)]_*$}
			&\multicolumn{1}{c}{$\mathcal{P}_{n,\ell}(t)$}&\multicolumn{1}{c|}{$[\mathcal{P}_{n,\ell}(t)]_*$}
			&\multicolumn{1}{c}{$\mathcal{P}_{n,\ell}(t)$}&\multicolumn{1}{c}{$[\mathcal{P}_{n,\ell}(t)]_*$}\\\hline
			1&0 &0.5&0.00996&0.5&0.01438&0.5&0.26481&0.5&0.33718&&0.55073&0.5&0.00882\\
		&&6.04037&15.41675&7.69908&19.04978&12.81262&20.19775&6.48812&9.33713&&4.15193&5.58510&14.39478\\
		2&0 &0.5 &0.07422&0.5&0.07017&0.5&0.00250&0.5&0.02601&0.5&0.20896&0.5&0.07541\\
		& &  4.05687&7.64390&5.20392&9.83466&8.98207&20.19775&4.49681&9.33713&2.54671&4.15193&3.74334&7.04773\\
		2&1 &0.5 &0.07422&0.5&0.07017&0.5&0.00248&0.5&0.02594&0.5&0.20837&0.5&0.07541\\
		& & 4.05683&7.64382&5.20386&9.83453&8.98105&20.19775&4.49604&9.33713&2.54509&4.15193&3.74331&7.04766\\
		3&0 &0.5 &0.08953&0.5&0.08946&0.5&0.03267&0.5&0.01398&0.5&0.04601&0.5&0.08946\\
		& & 3.17445&6.00121&4.08594&7.70278&7.36502&25.51282&3.70108&8.08915&2.08774&4.15193&2.92602&5.53707\\
		3&1 &0.5&0.08953&0.5&0.08945&0.5&0.03269&0.5&0.01401&0.5&0.04566&0.5&0.08946\\
		& &  3.17442&6.00116&4.08589&7.70269&7.36413&25.51564&3.70042&8.08530&2.08652&4.15193&2.92599&5.53702\\
		3&2 &0.5 &0.08953&0.5&0.08945&0.5&0.03274&0.5&0.014103&0.5&0.04498&0.5&0.08946\\
		& &   3.17435&6.00104&4.08579&7.70251&7.36235&25.52130&3.69910&8.07766&2.08409&4.15193&2.92594&5.53692\\\hline
			\hline
	\end{tabular}}
	\centering\caption{\label{table3.probability} Table of survival probability $P(t)=\frac{1}{2}$ and minimum value of survival probability of molecules in a TD moving boundary problem and the corresponding time.}
\end{table}
	\begin{figure}[h]
		\centering
		\includegraphics[width=19cm,height=12cm]{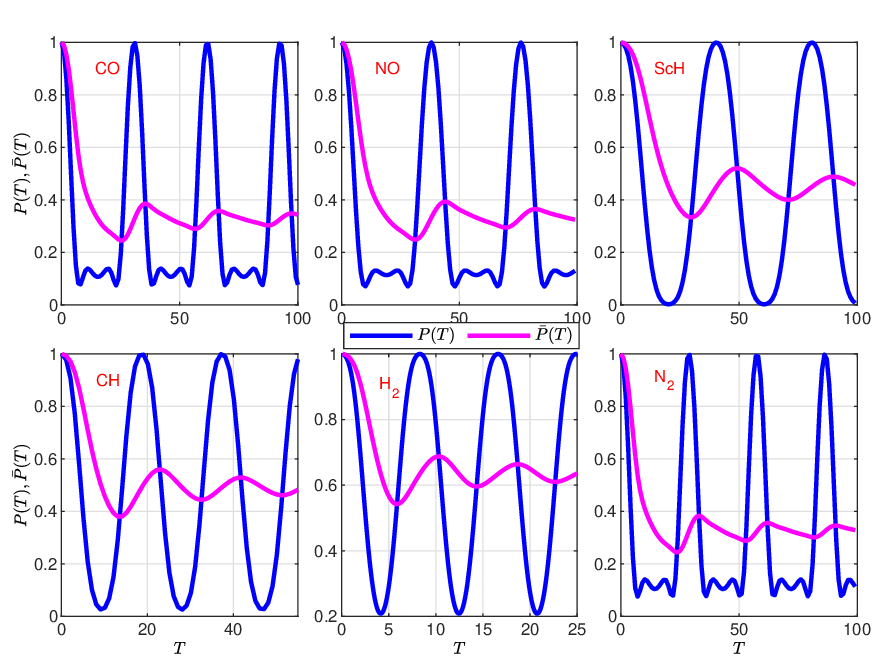}
		\caption{\label{fig.probability} Comparison of survival probability with average life-time, for six molecules in a moving boundary problem.}
	\end{figure}
	
\subsection{Survival probability and average life-time}\label{survival}
The auto-correlation function is an important property of a dynamical system. Using auto-correlation function one can define survival probability of a molecule in a TD quantum system. The survival probability represents the transition between initial and final states of a TD Hamiltonian. In most of the cases time evolution is obtained numerically for free boundary quantum system \cite{pfeifer1993,luo2005,boykin2007,khalil2015}. Survival probability is defined at a time t and it is periodic in time. It is used for finding the time energy uncertainty relation. If the survival probability is monotone decreasing then it is used to find the evolution time. But for non-decreasing function of time it is not suitable for evolution time. So, in this case average life-time is defined over a time domain $(0,t)$. The time evolution under a TD Hamiltonian with moving boundary condition is studied in this article first time.
The survival probability of an arbitrary state of $i$th molecule is obtained from the auto-correlation function as \cite{pfeifer1993,luo2005,boykin2007,khalil2015}:
\beq
\mathcal{P}_{n,\ell}(t)=|\mathcal{C}^{(i)}_{n,\ell}(0,t)|^2.
\eeq
Initially, $\mathcal{P}_{n,\ell}(0)=1$ and $0\le \mathcal{P}_{n,\ell}(t)\le1$ for $t\ge 0$. It is a periodic function of time. We observe that $\mathcal{P}_{n,\ell}(\frac{k\pi}{\om_1})=1$, where $k=0,1,2,\cdots$, and the minimum value of $[\mathcal{P}_{n,\ell}(t)]_*$ for six molecules are offered in Table \ref{table3.probability}. 
We observe that $\mathcal{P}_{n,\ell}(t)>0$ for $t\ge 0$ and $\mathcal{P}_{n,\ell}(t)$ never becomes zero. Therefore, $[\mathcal{P}_{n,\ell}(t)]_*\le \mathcal{P}_{n,\ell}(t)\le 1$ for $t\ge0$. 

Within the first periodic domain $[0,\frac{\pi}{\om_1}]$, we obtain $[\mathcal{P}_{n,\ell}]^*=P(0)=P(\frac{\pi}{\om_1})=1$ and the survival probability lies between $[\mathcal{P}_{n,\ell}(t)]_*$ and $1$ for $t\in(0,\frac{\pi}{\om_1})$ except at time $\frac{\pi}{2\om_1}$. 

Now, suppose $\mathcal{P}_{n,\ell}(t_1)=\eps$, then we observe that $\mathcal{P}_{n,\ell}(t_1)\le \frac{1}{2}\left(1+\sqrt{\mathcal{P}_{n,\ell}(2t_1)}\right)$, where $0<t_1<2t_1<\frac{\pi}{\om_1}$.

In Table \ref{table3.probability} we have shown the time $t^{(1/2)}$ for which the survival probability is $\mathcal{P}_{n,\ell}(t^{(1/2)})=\frac{1}{2}$ and it is observed that $\mathcal{P}_{n,\ell}(2t^{(1/2)})>0$. At $t=\frac{k\pi}{\om_1}$, for $k=1,2,3,\cdots$ the molecule behaves like that in the initial state. 

In Fig. \ref{fig.probability}, we have plotted the survival probability $\mathcal{P}_{n,\ell}(t)$ of six molecules in a confined pseudo-harmonic oscillator of TD moving boundary. 
Due to non monotone decreasing nature of the probability, we have defined the average life-time $\mathcal{\widetilde{P}}_{n,\ell}(t)$ as:
\beq 
\mathcal{\widetilde{P}}_{n,\ell}(t)=\ds\frac{1}{t}\int_0^t\mathcal{P}_{n,\ell}(t')dt',~t\ge0,
\eeq 
where $\mathcal{\widetilde{P}}_{n,\ell}(0)=\mathcal{P}_{n,\ell}(0)=1$. Then the average life-time decays from initial time, and it is oscillating but not periodic. Its length of oscillation decreases as time increases. It is found that $\mathcal{\widetilde{P}}_{n,\ell}(t)<1$ for $t>0$. Now, suppose $0\le \eps\le 1$. Then the evolution time $t^{(\eps)}$ is the minimum time for which average life-time is $\mathcal{\widetilde{P}}_{n,\ell}(t^{(\eps)})=\eps$. For a fixed value of $\epsilon$, it is difficult to find the exact value of $t^{(\eps)}$. In this paper we have computed the numerical value of $t^{(1/2)}$ and $t_*$ for which $\mathcal{\widetilde{P}}_{n,\ell}(t^{1/2})=1/2$ and $\mathcal{\widetilde{P}}_{n,\ell}(t_*)=[\mathcal{\widetilde{P}}_{n,\ell}(t)]_*$ is the minimum value of the average life-time. These are provided in Table \ref{table3.average-probability} for six molecules for some selected quantum states. In Fig. \ref{fig.probability} we display the average life-time of six molecules.

\subsection{Quantum similarity, disequilibrium and quantum similarity index}
The QSM between two states, having densities $\rho^{(i)}_{n_1,\ell_1,m_1}({\bf r},t)$ and $\rho^{(j)}_{n_2,\ell_2,m_2}({\bf r},t)$, of $i$th and $j$th molecules is defined by, 
\beq
\ba{ll}
\ds \mathcal{M}^{(i,j)}_{(n_1,\ell_1,m_1),(n_2,\ell_2,m_2)}(t)&=\ds\int \left[\rho^{(i)}_{n_1,\ell_1,m_1}({\bf r},t)\rho^{(j)}_{n_2,\ell_2,m_2}({\bf r},t)\right]d{\bf r}=M^{(r,i,j)}_{(n_1,\ell_1),(n_2,\ell_2)}(t)M_{(\ell_1,m_1),(\ell_2,m_2)}^{(\Om)},
\ea
\eeq 
where
\beq
\ds \mathcal{M}^{(r,i,j)}_{(n_1,\ell_1),(n_2,\ell_2)}(t)=\ds\int_{0}^{\widetilde{\b}^{(i,j)}_{n_1,n_2}} d^{(c,i)}_{n_1,\ell_1}(r,t)d^{(c,j)}_{n_2,\ell_2}(r,t)\,r^2dr,
\eeq
and
\beq
\ds \mathcal{M}_{(\ell_1,m_1),(\ell_2,m_2)}^{(\Om)}=\ds\int_{\phi=0}^{2\pi}\int_{\theta=0}^{\pi}\Om_{\ell_1,m_1}(\theta,\phi)\Om_{\ell_2,m_2}(\theta,\phi)\sin\theta\,d\theta\,d\phi,
\eeq
are respectively radial and angular components of quantum similarities, whereas $[0,\widetilde{\b}^{(i,j)}_{n_1,n_2}]$ is the common domain of the density functions $d^{(c,i)}_{n_1,\ell_1}(r,t)$ and $d^{(c,j)}_{n_2,\ell_2}(r,t)$. Since the TD radius of the confined spherical box is a function of spectroscopic parameters, $\widetilde{\beta}^{(i,j)}_{n_1,n_2}(T)=\min\{r^{(i)}_{0,(n_1,\ell_1)}\b^{(i)}(T),r^{(j)}_{0,(n_2,\ell_2)}\b^{(j)}(T)\}$ is used as an upper limit of the overlap integral. 
	
The quantity $\mathcal{M}^{(i,i)}_{(n,\ell,m),(n,\ell,m)}(T)$ characterizes disequilibrium in a given state $\psi^{(i)}_{n,\ell,m}({\bf r},t)$, which is denoted by $\mathcal{D}^{(i)}_{(n,\ell,m)}(t)$. Also it is called self-similarity \cite{carbo1980,gqsi} of the state $\psi^{(i)}_{n,\ell,m}({\bf r},t)$. Furthermore, it can also be written as a product of radial  $\left(\mathcal{R}^{(i)}_{(n,\ell)}(T)=\mathcal{M}^{(r,i,i)}_{(n,\ell),(n,\ell)}(T)\right)$ and angular components of disequilibrium $\left(\mathcal{A}_{(\ell,m)}=\mathcal{M}_{(\ell,m),(\ell,m)}^{(\Om)}\right)$. The radial disequilibrium $\mathcal{R}^{(i)}_{(n,\ell)}(t)$ of the state $\psi^{(i)}_{n,\ell,m}$ can be expressed \cite{jsd2010} analytically as, 
\beq
\ds\mathcal{R}^{(i)}_{(n,\ell)}(t)=\ds\frac{12[r_{0,(n,\ell)}^{(i)}]^3[N_n^{(i)}]^4}{[\b^{(i)}]^3}\sum\limits_{k=0}^{4n}\frac{B_{4+k,4}\left(c_{n,0}^{(\ell,i)},2!c_{n,1}^{(\ell,i)},\cdots,(k+1)!c_{n,k}^{(\ell,i)}\right)\left[\G\left(k+2L^{(i)}+\frac{3}{2}\right)-\G\left(k+2L^{(i)}+\frac{3}{2},2\lambda_{n,\ell}^{(i)}\right)\right]}{(2\lambda_{n,\ell}^{(i)})^{2L+\frac{3}{2}}(4+k)!2^k},
\eeq
where 
\beq
\ds c_{n,k}^{(\ell,i)}=\ds\left\{\ba{ll}\ds\frac{(-n)_k}{(L^{(i)}(\ell)+\frac{3}{2})_k\,k!},&~ \mbox{if}~ k\le n\\
0,&~\mbox{if}~ k>n,\ea \right\}, 
\eeq 
and $\lambda_{n,\ell}^{(i)}$ is the largest positive root of ${}_1F_1\left[-n,L^{(i)}+\frac{3}{2},\lambda_{n,\ell}^{(i)}\right]=0$. It may be noted that $\mathcal{R}^{(i)}_{(n,\ell)}(t)$ is time dependent; also it is a function of $a,b,c_0,r_{0,(n,\ell)},t_0$ which are associated with the moving boundary wall with TD IDR $r_{0,{n,\ell}}^{(i)}\b^{(i)}(T)$ and the constraint in Eq.~(\ref{abc0t0}). The angular disequilibrium $\mathcal{A}_{(\ell,m)}$, on the other hand, is given by \cite{jsd.epjd2009,3jsymbols}, 
\beq
\mathcal{A}_{(\ell,m)}=\ds\frac{(2\ell+1)^2}{4\pi}\ds\sum\limits_{j=0}^{2\ell}(2j+1)\left(\ba{ccc} \ell & \ell & j\\0 & 0 & 0 \ea \right)^2\left(\ba{ccc} \ell & \ell & j\\m & m & -2m \ea \right)^2,
\eeq  
where $\left(\ba{ccc} a & b & c\\ \a & \b & \g \ea \right)$ is called the $3j$ symbol. Then disequilibrium of $\psi^{(i)}_{n,\ell,m}$ can be expressed as: $\mathcal{D}^{(i)}_{(n,\ell,m)}(t)=\mathcal{R}^{(i)}_{(n,\ell)}(t)\mathcal{A}_{(\ell,m)}$.

	\begin{table}[]
	\centering
	\scalebox{.78}{\begin{tabular}{ll|ll|ll|ll|ll|ll|ll}\hline\hline
			$n$&$\ell$ & \multicolumn{2}{c|}{CO} & \multicolumn{2}{c|}{NO}&  \multicolumn{2}{c|}{ScH}& \multicolumn{2}{c|}{CH}& \multicolumn{2}{c|}{H$_2$}& \multicolumn{2}{c}{N$_2$}\\ \hline
			&
			&\multicolumn{1}{c}{$\mathcal{\widetilde{P}}_{n,\ell}(t^{(\frac{1}{2})})$}&\multicolumn{1}{c|}{$[\mathcal{\widetilde{P}}_{n,\ell}(t)]_*$} &\multicolumn{1}{c}{$\mathcal{\widetilde{P}}_{n,\ell}(t^{(\frac{1}{2})})$}&\multicolumn{1}{c|}{$[\mathcal{\widetilde{P}}_{n,\ell}(t)]_*$}
			&\multicolumn{1}{c}{$\mathcal{\widetilde{P}}_{n,\ell}(t^{(\frac{1}{2})})$}&\multicolumn{1}{c|}{$[\mathcal{\widetilde{P}}_{n,\ell}(t)]_*$}
			&\multicolumn{1}{c}{$\mathcal{\widetilde{P}}_{n,\ell}(t^{(\frac{1}{2})})$}&\multicolumn{1}{c|}{$[\mathcal{\widetilde{P}}_{n,\ell}(t)]_*$}
			&\multicolumn{1}{c}{$\mathcal{\widetilde{P}}_{n,\ell}(t^{(\frac{1}{2})})$}&\multicolumn{1}{c|}{$[\mathcal{\widetilde{P}}_{n,\ell}(t)]_*$}
			&\multicolumn{1}{c}{$\mathcal{\widetilde{P}}_{n,\ell}(t^{(\frac{1}{2})})$}&\multicolumn{1}{c}{$[\mathcal{\widetilde{P}}_{n,\ell}(t)]_*$}\\\hline
			1&0 & 0.5&0.2907& 0.5&0.3045 &&0.5734&&0.6245&&0.7589&0.5&0.2864\\
& &   12.3648&23.350&15.8030 &28.694& &28.690&&13.181&&5.78&11.4241&21.840\\\hline
2&0 &0.5 &0.2450 &0.5&0.2486   &0.5&0.3340&0.5&0.3788&&0.5428&0.5&0.2437\\
& &   8.3854&25.437&10.7296 &31.224&18.1509& 29.757  &9.1796&13.578&&5.881&7.7431&23.799\\\hline
2&1 & 0.5&0.2451&0.5&0.2487&0.5&0.3340&0.5&0.3787&&0.5424&0.5&0.2437\\
& &   8.3853&25.437&10.7295&31.224&18.1487&29.758&9.1777&13.578&&5.881&7.7431&23.799\\\hline
3&0 & 0.5&0.2024&0.5&0.2137&0.5&0.2742&0.5&0.2905&0.5&0.4055&0.5&0.1990\\
& &   6.6349&26.281& 8.5380&32.340&14.9348&31.031&7.4718&14.063&4.2945&5.995&6.1157&24.5718\\\hline
3&1 &0.5 &0.2024&0.5&0.2137&0.5&0.2742&0.5&0.2905&0.5&0.4051&0.5&0.1990\\
& &  6.6348&26.281&8.5379&32.340&14.9332&31.032&7.4705&14.063&4.2912&5.996&6.1157&24.5718\\\hline
3&2 &0.5 &0.2024&0.5&0.2136&0.5&0.2742&0.5&0.2903&0.5&0.4043&0.5&0.1990\\
& &   6.6347&26.281&8.5377&32.340&14.9298&31.034&7.4680&14.0656&4.2848&5.9969&6.1156&24.5718\\\hline
			\hline
	\end{tabular}}
	\centering\caption{\label{table3.average-probability} Table of time evolution $t^{(1/2)}$ of average life-time $\widetilde{P}(t^{(1/2)})=\frac{1}{2}$ and global minimum value of average life-time $\widetilde{P}(t_{1*})$ for six molecules in a TD moving boundary problem and the corresponding time.}
\end{table}

Next, the QSI between two quantum states $\psi^{(i)}_{n_1,\ell_1,m_1}$ and $\psi^{(j)}_{n_2,\ell_2,m_2}$ is given by, 
\beq
\ds \mathcal{S}^{(i,j)}_{(n_1,\ell_1,m_1),(n_2,\ell_2,m_2)}(t)=\ds\frac{\mathcal{M}^{(i,j)}_{(n_1,\ell_1,m_1),(n_2,\ell_2,m_2)}(T)}{\sqrt{\mathcal{D}^{(i)}_{(n_1,\ell_1,m_1)}(T)}\sqrt{\mathcal{D}^{(j)}_{(n_2,\ell_2,m_2)}(T)}}.
\eeq
It follows that, TD QSI is a function of $\ds\frac{r_{0,(n_1,\ell_1)}^{(i)}\,\beta^{(i)}(T)}{r_{0,(n_2,\ell_2)}^{(j)}\,\beta^{(j)}(T)}$, which is a ratio of two internuclear distances of $i$th and $j$th molecules having quantum numbers $(n_1,\ell_1,m)$ and $(n_2,\ell_2,m)$, at the same time. 
	
\begin{figure}[t]
\centering
\includegraphics[width=19cm,height=12cm]{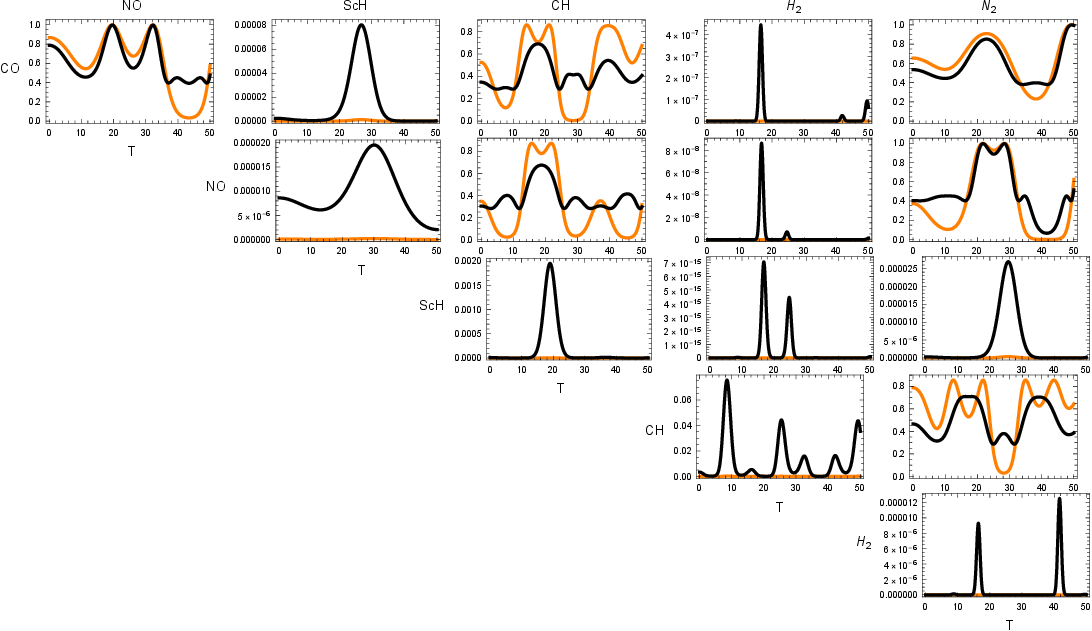}
\caption{\label{fig5.qsi} Comparison of TD QSI between pairs of quantum states of diatomic molecules, for $b=0.1$ and $t_0=1973.29\,sec$. The orange and black lines represent $n=1,\ell=0$ and $n=2,\ell=1$ respectively.}
\end{figure}

We have also calculated $\mathcal{S}^{(i,j)}_{(n_1,\ell,m),(n_2,\ell,m)}(t)$ of a pair of molecules having same angular and azimuthal quantum numbers, which is independent of the quantum number $m$; this is denoted as $\mathcal{S}^{(i,j)}_{(n_1,n_2,\ell)}(t)$. Also one can observe that $\mathcal{S}^{(i,j)}_{(n_1,n_2,\ell)}(t)$ is a QSI of two radial density functions. We have also calculated the radial QSI among some pairs of molecules. For $i$ and $j$ molecules having quantum numbers $(n_1,\ell,m)$ and $(n_2,\ell,m)$, the QSI $\mathcal{S}^{(i,j)}_{(n_1,n_2,\ell)}(t)$ is given by, 
\beq
\mathcal{S}^{(i,j)}_{(n_1,n_2,\ell)}(t)=\ds\frac{\mathcal{M}^{(i,j)}_{(n_1,\ell),(n_2,\ell)}(T)}{\sqrt{\mathcal{R}^{(i)}_{(n_1,\ell)}(T)}\sqrt{\mathcal{R}^{(j)}_{(n_2,\ell)}(T)}}.
\eeq 
The TD IDR between two molecular states having quantum numbers $(n_1,\ell,m)$ and $(n_2,\ell,m)$ can be written as: $\ds IDR^{(i,j)}_{n_1,n_2,\ell}(T,T)=\ds\frac{r_{0,(n_1,\ell)}^{(i)}\beta^{(i)}(T)}{r_{0,(n_2,\ell)}^{(j)}\beta^{(j)}(T)}$. After some mathematical manipulations, we find the analytical form of QSI in terms of $IDR^{(i,j)}_{n_1,n_2,\ell}(T)$ as, 
\beq
\mathcal{S}^{(i,j)}_{(n_1,n_2,\ell)}(T)=\ds\left(IDR^{(i,j)}_{n_1,n_2,\ell}(T,T)\right)^{2L^{(i)}+\frac{3}{2}}\frac{(2\lambda_{n_1,\ell}^{(i)})^{L^{(i)}+\frac{3}{4}}(2\lambda_{n_2,\ell}^{(j)})^{L^{(j)}+\frac{3}{4}}}{(\lambda_{n_1,\ell}^{(i)}+\widetilde{\lambda}_{n_1,n_2,\ell}^{(i,j)}(T))^{L^{(i)}+L^{(j)}+\frac{3}{2}}}\frac{\widetilde{\mathcal{J}}^{(i,j)}_{n_1,n_2,\ell}\left[x^{(i,j)}_{n_1,n_2,\ell}(T)\right]}{\sqrt{\mathcal{J}_{(n_1,\ell)}^{(i)}(2\lambda_{n_1,\ell}^{(i)}})\sqrt{\mathcal{J}_{(n_2,\ell)}^{(j)}(2\lambda_{n_2,\ell}^{(j)})}},
\eeq 
where $\widetilde{\mathcal{J}}^{(i,j)}_{n_1,n_2,\ell}(x)$ is given as follows, 
\beq
\ba{ll}
\ds\widetilde{\mathcal{J}}^{(i,j)}_{n_1,n_2,\ell}(x)&=\ds\int_0^{x}e^{-z}z^{L^{(i)}+L^{(j)}+\frac{1}{2}}\left[{}_1F_1\left(-n_1,L^{(i)}+\frac{3}{2};\frac{\lambda_{n_1,\ell}^{(i)}\,z}{\lambda_{n_1,\ell}^{(i)}+\widetilde{\lambda}_{n_1,n_2,\ell}^{(i,j)}(T)}\right){}_1F_1\left(-n_2,L^{(j)}+\frac{3}{2};\frac{\widetilde{\lambda}_{n_1,n_2,\ell}^{(i,j)}(T)\,z}{\lambda_{n_1,\ell}^{(i)}+\widetilde{\lambda}_{n_1,n_2,\ell}^{(i,j)}(T)}\right)\right]^2\,dz\\\\
&=\ds\sum\limits_{k=0}^{2n_1}\sum\limits_{l=0}^{2n_2}\frac{4\,B_{2+k,2}\left(c_{n_1,0}^{(\ell,i)},2!c_{n_1,1}^{(\ell,i)},\cdots,(k+1)!c_{n_1,k}^{(\ell,i)}\right)}{(2+k)!}\frac{B_{2+l,2}\left(c_{n_2,0}^{(\ell,j)},2!c_{n_2,1}^{(\ell,j)},\cdots,(l+1)!c_{n_2,l}^{(\ell,j)}\right)}{(2+l)!}\left(\frac{\lambda_{n_1,\ell}^{(i)}}{\lambda_{n_1,\ell}^{(i)}+\widetilde{\lambda}_{n_1,n_2,\ell}^{(i,j)}(T)}\right)^k\\\\
&\quad ~\ds\times\left(\frac{\widetilde{\lambda}_{n_1,n_2,\ell}^{(i,j)}}{\lambda_{n_1,\ell}^{(i)}+\widetilde{\lambda}_{n_1,n_2,\ell}^{(i,j)}(T)}\right)^l \left[\G\left(L^{(i)}+L^{(j)}+k+l+\frac{3}{2}\right)-\G\left(L^{(i)}+L^{(j)}+k+l+\frac{3}{2},x\right)\right],
\ea 
\eeq
and $\ds\mathcal{J}_{(n_1,\ell)}^{(i)}(2\lambda_{n_1,\ell}^{(i)})$, $\mathcal{J}_{(n_2,\ell)}^{(j)}(2\lambda_{n_2,\ell}^{(j)})$ are obtained from the following integral, 
\beq
\ba{ll}
\ds\mathcal{J}_{(n,\ell)}^{(i)}(x)&=\ds\int_0^{x}e^{-z}z^{2L^{(i)}+\frac{1}{2}}\left[_{1}F_1\left(-n,,L^{(i)}+\frac{3}{2},\frac{z}{2}\right)\right]^4\,dz\\\\
&=\ds\sum\limits_{k=0}^{4n}\frac{24\,B_{4+k,4}\left(c_{n,0}^{(\ell,i)},2!c_{n,1}^{(\ell,i)},\cdots,(k+1)!c_{n,k}^{(\ell,i)}\right)\left[\G\left(k+2L^{(i)}+\frac{3}{2}\right)-\G\left(k+2L^{(i)}+\frac{3}{2},x\right)\right]}{(4+k)!2^k},
\ea 
\eeq
where
\beq
\ba{ll}
\ds\widetilde{\lambda}_{n_1,n_2,\ell}^{(i,j)}(T)=\ds\lambda_{n_2,\ell}^{(j)}
\left(IDR^{(i,j)}_{n_1,n_2,\ell}(T,T)\right)^2,~
\ds x^{(i,j)}_{n_1,n_2,\ell}(T)=\ds\min\left\{\lambda_{n_1,\ell}^{(i)}+\widetilde{\lambda}_{n_1,n_2,\ell}^{(i,j)}(T),\frac{\lambda_{n_1,\ell}^{(i)}+\widetilde{\lambda}_{n_1,n_2,\ell}^{(i,j)}(T)}{\left[IDR^{(i,j)}_{n_1,n_2,\ell}(T,T)\right]^2}\right\}.
\ea 
\eeq
If $b=0$, then the TD QSI reduces to simple QSI in a confined quantum system in an infinite spherical well of finite radius. The calculated QSI, $\mathcal{S}^{(i,j)}_{(n,\ell)}$ between the pairs of molecules are plotted in Fig.~\ref{fig5.qsi} for $b=0.1, t_0=1000$. It is seen that the QSIs fluctuate with time. The QSI at an initial time and the numerically calculated values of local maxima of QSI, $\mathcal{S}^{(i,j)}_{(n,\ell)}(t^*)$ at $t^*$ are reported in Table~\ref{table5.qsi}. 

\begin{table}[t] 
\centering
\scalebox{.9}{\begin{tabular}{l|llllll|c}\hline\hline
	$i\,\searrow\,j$& \multicolumn{1}{c}{j=CO} & \multicolumn{1}{c}{j=NO}&  \multicolumn{1}{c}{j=ScH}& \multicolumn{1}{c}{j=CH}& \multicolumn{1}{c}{j=H$_2$}& \multicolumn{1}{c|}{j=N$_2$}&$\mathcal{S}^{(i,j)}_{(n,\ell)}(T)$\\\hline
	CO&1.0&0.868333952&2.423669904(-8)&0.524496109&1.747212557(-25)&0.654920461&b=0\\
	CO&1.0&0.868333952&2.423669904(-8)&0.524496109&1.747212557(-25)&0.654920461&$\mathcal{S}^{(CO,j)}_{(n,\ell)}(0)$\\
	CO&1.0&0.999436139&0.000001255&0.859423881 &3.175936271(-16)&0.909617070&$\mathcal{S}^{(CO,j)}_{(n,\ell)}(T^*)$\\
	CO&1.0&19.548160&26.78655&14.15627&16.51177&22.8471&$T^*$\\\hline
	
	NO&&1.0&8.890869474(-8)&0.3470473113&2.203966332(-25)&0.37296019&b=0\\
	NO&&1.0&8.890869474(-8)&0.3470473113&2.203966332(-25)&0.37296019&$\mathcal{S}^{(NO,j)}_{(n,\ell)}(0)$\\
	NO&&1.0&2.235586880(-7)&0.872561122&9.565862223(-17)&0.99902650&$\mathcal{S}^{(NO,j)}_{(n,\ell)}(T^*)$\\
	NO&&1.0&30.08924&15.7241&16.737733&21.6771&$T^*$\\\hline
	
	ScH&&&1.0&5.91103595(-9)&4.439267281(-24)&2.756610896(-9) &b=0\\
	ScH&&&1.0&5.91103595(-9)&4.439267281(-24)&2.756610896(-9)&$\mathcal{S}^{(ScH,j)}_{(n,\ell)}(0)$\\
	ScH&&&1.0&3.741847821(-6)&2.419859864(-20)&3.757717056(-7)&$\mathcal{S}^{(ScH,j)}_{(n,\ell)}(T^*)$\\
	ScH&&&1.0&18.9879&16.773936&25.71519&$T^*$\\\hline
	
	CH&&&&1.0&2.84147921(-6)&0.7878290354&b=0\\
	CH&&&&1.0&2.84147921(-6)&0.7878290354&$\mathcal{S}^{(CH,j)}_{(n,\ell)}(0)$\\
	CH&&&&1.0&0.000318282&0.855235891&$\mathcal{S}^{(CH,j)}_{(n,\ell)}(T^*)$\\			
	CH&&&&1.0&8.502434&12.49156&$T^*$\\\hline
	
	H$_ 2$&&&&&1.0&5.725709599(-23)&b=0\\
	H$_ 2$&&&&&1.0&5.725709599(-23)&$\mathcal{S}^{(H_2,N_2)}_{(n,\ell)}(0)$\\
	H$_ 2$&&&&&1.0&2.221070777(-14)&$\mathcal{S}^{(H_2,N_2)}_{(n,\ell)}(T^*)$\\
	H$_ 2$&&&&&1.0&16.411008&$T^*$\\\hline\hline
\end{tabular}}
\centering\caption{\label{table5.qsi} Comparison of QSI $\ds\mathcal{S}^{(i,j)}_{(n,\ell,m)}(t)$ among diatomic molecules for $i,j\in $ \{CO,NO,ScH,CH,H$_2$,N$_2$\} for $n=2,\ell=1$, $t_0=1973.29$ sec, $b=0.1$ at initial and a local maximum time. $\ds\mathcal{S}^{(i,j)}_{(n,\ell)}(0)$ and $\ds\mathcal{S}^{(i,j)}_{(n,\ell)}(T^*)$ represent the initial and local maximum value of QSI at $T^*$. In each segment first row ($b=0$) shows QSI in a time-independent confined system.}
\end{table}

\section{Conclusion}\label{sec5.con}
Exact solution of TDSE with a confined pseudoharmonic oscillator is presented for arbitrary angular quantum number. The density function and internuclear distance of molecules depend on the solution of Ermakov equation. The shorter and longer range of internuclear distances for molecules have been obtained analytically. The estimated numerical values have been presented in tabular form. The minimum value $\hbar(\Delta s)_{n,\ell}\sqrt{\mathcal{J}_{n,\ell}-\left\langle 1/s\right\rangle_{n,\ell,m}}$ of TD Heisenberg uncertainty product is always greater than the minimum uncertainty product $\hbar/2$. The average energies are real, whose closed form expressions are provided, along with the corresponding numerical values. The latter are compared with the ro-vibrational energies. The average energy of the system over a long time collapses to a fixed energy value. Analytical form of average force and average pressure also are defined for non-interacting particles confined in a spherical box of TD radius. The optimum values of the average force and average pressure are defined analytically. The corresponding numerical values are shown in tabular form for six molecules Then time correlation functions such as cross and auto correlations are analytically obtained and graphically shown for selected molecules. The property of survival probability at two different times of a quantum state is investigated. The minimum value of survival probability and first half-lifetime of molecules are reported. The survival probability of a confined particle is periodic in time and some times it is equal to one, which shows that the particle goes back to the initial state. To find the evolution time of the initial state we have defined the average life-time of molecules. We observe that it is never zero. Then expressions for QSM, disequilibrium and QSI have been provided in closed form. Results are given in tabular and graphical form. Then local maximum values of QSI and initial values of QSI are offered. It appears that, IDR at two different time domain appears to be the most important quantity for studying the time correlation function. It depends on $n$ and $\ell$ quantum numbers. The IDR at the same time $T$ fluctuates with respect to $T$ whereas at different time domain $(t_1, t_2)$, it behaves as a periodic function of $t_1$ for fixed $t_2$, and of $t_2$ for fixed $t_1$. If $t_0$ has a large value then $c_0$ must be large and in this case $\beta$ is constant and the TD moving boundary problem reduces to a fixed boundary, which is the time independent confined system.

\section*{Acknowledgement}
DN gratefully acknowledges financial support from TARE, DST-SERB, New Delhi (sanction order: TAR/2021/000142). AKR acknowledges funding support from DST SERB (sanction order CRG/2023/004463). AH gratefully acknowledges JRF financial support from CSIR, New Delhi (09/0921(16264)/2023-EMR-I). 
	
\section*{Author Contribution}
\noindent All authors have been contributed equally to this work.
\section*{Conflict of interest} 
The authors declare that they have no conflict of interest.
\section*{Data availability statement}
All data generated or analyzed during this study are included in this article.


\begin{thebibliography}{}
	\bibitem{campoy2002} G.~Campoy, N.~Aquino and V.~Granados, \textit{Energy eigenvalues and Einstein coefficients for the one-dimensional confined harmonic oscillators}, J.~Phys.~A \textbf{35}, 4903 (2002). 
	\bibitem{sen2006} K.~D.~Sen and A.~K.~Roy, \textit{Studies on the 3D confined potentials using generalized pseudospectral approach}, Phys.~Lett.~A \textbf{357}, 112 (2006). 
	\bibitem{stevanovic2008} Lj.~Stevanovi\'c and K.~D.~Sen, \textit{Eigenspectrum properties of the confined 3D harmonic oscillator}, J.~Phys.~A \textbf{41}, 225002 (2008).  
	\bibitem{montgomery2010} H.~E.~Montgomery Jr., G.~Campoy and N.~Aquino, \textit{The confined N-dimensional harmonic oscillator revisited}, Phys.~Scr.~ \textbf{81}, 045010 (2010). 
	\bibitem{mukherjee2019} N.~Mukherjee and A.~K.~Roy, \textit{Information analysis in free and confined harmonic oscillators}, in \textit{Harmonic oscillators: types, functions and applications}, edited by Yilun Shang, pp.~1-86, (Nova Science Publishers, Hauppauge, NY), (2019). 
	\bibitem{burrows2006} B.~L.~Burrows and M.~Cohen, \textit{Exact solutions for spherically confined hydrogen-like atoms}, Int.~J.~Quant.~Chem. \textbf{106}, 478 (2006). 
	\bibitem{ciftci2009} H.~Ciftci, R.~L.~Hall and N.~Saad, \textit{Study of a confined hydrogen-like atom by the asymptotic iteration method}, Int.~J.~Quant.~Chem. \textbf{109}, 931 (2009). 
	\bibitem{roy2015} A.~K.~Roy, \textit{Spherical confinement of Coulombic systems inside an impenetrable box: H atom and the H\'ulthen potential}, Int.~J.~Quant.~Chem. \textbf{115}, 937 (2015). 
	\bibitem{canic2020} S.~\v{C}ani\'c, \textit{Moving boundary problems}, Bull.~Am.~Math.~Soc. \textbf{58}, 79 (2020).
	\bibitem{yuce2004} C.~Y\"uce, \textit{Exact solvability of moving boundary problems}, Phys.~Lett.~A \textbf{327}, 107 (2004). 
	\bibitem{munier1981} A.~Munier, J.~R.~Burgan, M.~Felix and E.~Fijalkow, \textit{Schr\"odinger equation with time-dependent boundary conditions}, J.~Math.~Phys.~ \textbf{22}, 1219 (1981).  
	\bibitem{cervero1999} J.~M.~Cerver\'o and J.~D.~Lejarreta, \textit{The time-dependent canonical formalism: Generalized harmonic oscillator and the infinite square well with a moving boundary}, Europhys.~Lett. \textbf{45}, 6 (1999). 
	\bibitem{centeno2020} M.~Ahumada-Centeno, P.~Amore, F.~M.~Fern\'andez and J.~Manzanares-Martinez, \textit{Quantum particles in a moving potential}, Phys.~Scr.~ \textbf{95}, 065405 (2020). 
	\bibitem{li2001} L.~Li and B.~Z.~Li, \textit{Exact evolving states for a class of generalized time-dependent quantum harmonic oscillators with a moving boundary}, Phys.~Lett.~A \textbf{291}, 190 (2001). 
	\bibitem{lejarreta1999} J.~D.~Lejaretta, \textit{The generalized harmonic oscillator and the infinite square well with a moving boundary}, J.~Phys.~A \textbf{32}, 4749 (1999). 
	\bibitem{rakhmanov2018} S.~Rakhmanov, D.~Matrasulov, and V.~I.~Matveev, \textit{Quantum dynamics of a hydrogen-like atom in a time-dependent box: Non-adiabatic regime}, Eur.~Phys.~J.~D \textbf{72}, 177 (2018).
	\bibitem{nieto2009} M.~L.~Glasser, J.~Mateo, J.~Negro and L.~M.~Nieto, \textit{Chaos Solitons \& Fractals} \textbf{41}, 2067 (2009).
	\bibitem{duffin2019} C.~Duffin and A.~G.~Dijkstra, \textit{Controlling a quantum system via its boundary conditions}, Eur.~Phys.~J.~D \textbf{73}, 221 (2019).
	\bibitem{nath2020} D.~Nath and P.~Roy, \textit{Time-dependent rationally extended P\"oschl-Teller potential and some of its properties}, Eur.~Phys.~J.~Plus {\bf 135}, 802 (2020).
	\bibitem{carbo-dorca2022} R.~Carb\'o-Dorca and D.~Nath, {\it Average energy and quantum similarity of a time dependent quantum system subject to P\"oschl-Teller potential}, J.~Math.~Chem. \textbf{60}, 440 (2022).
	\bibitem{nath2022} D.~Nath and R.~Carb\'o-Dorca, {\it Analysis of solutions of time-dependent Schr\"odinger equation of a particle trapped in a spherical box}, J.~Math.~Chem. \textbf{60}, 1089 (2022).
	\bibitem{nath2023jmc} D.~Nath and A.~K. Roy, \textit{Average energy and Shannon entropy of a confined harmonic oscillator in a time‑dependent moving boundary}, J.~Math.~Chem. \textbf{61}, 1491 (2023). 
	\bibitem{nath2021} D.~Nath, \textit{Comparison between time-independent and time-dependent quantum systems in the context of energy, Heisenberg uncertainty, average energy, force, average force and thermodynamic quantities}, arXiv:2110.05609v2 (2021).
	\bibitem{nath2022a} D.~Nath, \textit{Study of uncertainty, average energy, and thermodynamic quantities of the time dependent quantum system}, J.~Math.~Chem. \textbf{60}, 1819 (2022).		
	\bibitem{goldman1961} I.~I.~Goldman and V.~D.~Krivchenkov, \textit{Problems in Quantum Mechanics}, Pergamon Press, New York (1961).
	\bibitem{popov2001} D.~Popov, \textit{Barut-Girardello coherent states of the pseudoharmonic oscillator}, J.~Phys.~A \textbf{34}, 5283 (2001).
	\bibitem{ikhdair2007} S.~Ikhdair and R.~Sever, \textit{Exact polynomial eigensolutions of the Schr\"odinger equation for the pseudoharmonic potential}, J.~Mol.~Struct.: THEOCHEM, \textbf{806}, 155 (2007).
	\bibitem{ikhdair2007a} S.~M.~Ikhdair and R.~C.~Sever, \textit{Exact solutions of the radial Schr\"odinger equation for some physical potentials}, Cent.~Eur.~J.~Phys. \textbf{5}, 516 (2007). 
	\bibitem{sever2008} R.~Sever, C.~Tezcan, M.~Aktas and \"O. Yesiltas, \textit{Exact solution of Schr\"odinger equation for pseudoharmonic potential}, J.~Math.~Chem. \textbf{43}, 845 (2008).
	\bibitem{oyewumi2008} K.~J.~Oyewumi, F.~O.~Akinpelu and A.~D.~Agboola, \textit{Exactly complete solutions of the pseudoharmonic potential in N-dimensions}, Int.~J.~Theor.~Phys. \textbf{47}, 1039 (2008). 
	\bibitem{oyewumi2012} K.~J.~Oyewumi and K.~D.~Sen, \textit{Exact solutions of the Schr\"odinger equation for the pseudoharmonic potential: An application to some diatomic molecules}, J.~Math.~Chem. \textbf{50}, 1039 (2012).
	\bibitem{ikhdair2012} S.~M.~Ikhdair and M.~Hamzavi, \textit{A quantum pseudo dot system with two-dimensional pseudoharmonic oscillator in external magnetic and Aharonov-Bohm fields}, Physica B \textbf{407}, 4198 (2012).
	\bibitem{arda2012} A.~Arda and R.~Sever, \textit{Exact solutions of the Schr\"odinger equation via Laplace transform approach: Pseudoharmonic potential and Mie-type potentials}, J.~Math.~Chem. \textbf{50}, 971 (2012).
	\bibitem{akcay2012} H.~Akcay and R.~Sever, \textit{Analytical solutions of Schr\"odinger equation for the diatomic molecular potentials with any angular momentum}, J.~Math.~Chem. \textbf{50}, 1973 (2012).
	\bibitem{yahya2015} W.~A.~Yahya, K.~J.~Oyewumi and K.~D.~Sen, \textit{Position and momentum information-theoretic measures of the pseudoharmonic potential}, Int.~J.~Quant.~Chem. \textbf{115}, 1543 (2015).
	\bibitem{rani2018} R.~Rani, S.~B.~Bhardwaj and F.~Chand, \textit{Bound state solutions to the Schr\"dinger equation for some diatomic molecules}, Pramana J. Phys. \textbf{91}, 46 (2018).
	\bibitem{ghosh2020} P.~Ghosh and D.~Nath, \textit{Exact solutions and spectrum analysis of a non-central potential in the presence of vector potential}, Int.~J.~Quant.~Chem. \textbf{120}, e26153 (2020).
	
    \bibitem{atomic-molb} J.~F.~Ogilvie, \textit{The Vibrational and Rotational Spectrometry of Diatomic molecules}, Chap. IV, Academic Press, San Diago, CA (1998).	
	\bibitem{solution.beta} E.~Pinney, \textit{The nonlinear differential equation $y''+p(x)y+cy^{-3}=0$.}, Proc.~Amer.~Math.~Soc. \textbf{1}, 681 (1950). 		
	\bibitem{kds2007} H.~E.~Montgomery Jr., N.~A.~Aquino and K.~D.~Sen, \textit{Degeneracy of confined D-dimensional harmonic oscillator}, Int.~J.~Quantum.~Chem. {\bf 107}, 798 (2007).
	\bibitem{riordan1980} J.~Riordan, \textit{An Introduction to Combinatorial Analysis}, Princeton Univ. Press, New Jersey, (1980).
	\bibitem{nath2022_nld} D.~Nath and A.~K. Roy, \textit{Time-correlation function and average energy of molecules in presence of Deng-Fan potential in a moving boundary}, Nonlinear Dyn. \textbf{110}, 1597 (2022). 
	\bibitem{nakamura2011} K.~Nakamura, S.~K.~Avazbaev, Z.~A.~Sobirov, D.~U.~Matrasulov and T.~Monnai, \textit{Ideal quantum gas in an expanding cavity: Nature of non-adiabatic force}, Phys.~Rev.~E \textbf{83}, 041133 (2011).
	\bibitem{chen2010} X.~Chen, A.~Ruschhaupt, S.~Schmidt, A.~del~Campo, D.~Gu\'ery-Odelin, and J.~G.~Muga, \textit{Fast Optimal Frictionless Atom Cooling in Harmonic Traps: Shortcut to Adiabaticity}, Phys.~Rev.~Lett. \textbf{104}, 063002 (2010).
	\bibitem{mousavi2012} S.~V.~Mousavi, \textit{Quantum dynamics in a time-dependent hard-wall spherical trap}, Europhys. Lett. \textbf{99}, 30002 (2012).
	\bibitem{pfeifer1993} P.~Pfeifer, \textit{How fast can a quantum state change with time?}, Phys.~Rev.~Lett.  \textbf{70}, 3365 (1993).
	\bibitem{luo2005} S.~Luo, \textit{On survival probability of quantum states}, J.~Phys.~A: Math.~Gen. \textbf{38}, 2991 (2005).
	\bibitem{boykin2007} T.~B.~Boykin, N.~Kharche, and G.~Klimeck, \textit{Evolution time and energy uncertainty}, Eur.~J.~Phys. \textbf{28}, 673 (2007).
	\bibitem{khalil2015} T.~Khalil and J.~Richert, \textit{Excitation of time-dependent quantum systems: An application of time-energy uncertainty relations}, Physica B \textbf{457}, 78 (2015).
	
	\bibitem{tokmakoff2014} A.~Tokmakoff, \textit{Time-Dependent Quantum Mechanics and Spectroscopy}, The University of Chicago, Chicago, (2014).
	\bibitem{fring2020} A.~Fring and R.~Tenney, {\it Time-independent approximations for time-dependent optical potentials}, Eur.~Phys.~J.~Plus \textbf{135}, 163 (2020).
	
	
	\bibitem{carbo1980} M.~A.~R.~Carb\'o-Dorca and L.~Leyda, {\it How Similar is a Molecule to Another? An Electron Density Measure of Similarity between Two Molecular Structures}, Int.~J.~Quant~Chem. \textbf{17}, 1185 (1980).
	\bibitem{gqsi} P.~Ghosh and D. Nath, \textit{Generalized quantum similarity index: An application to pseudoharmonic oscillator with isospectral potentials in 3D}, Int.~J.~Quant.~Chem. \textbf{121}, e26517 (2021).
	
	\bibitem{jsd2010} P.~S\'anchez-Moreno, J.~S.~Dehesa, D.~Manzano and R.~J.~Y\'a\~nez, \textit{Spreading lengths of Hermite polynomials}, J.~Comput.~Appl.~Math. \textbf{233}, 2136 (2010).
	\bibitem{jsd.epjd2009} J.~S.~Dehesa, S.~Lopez-Rosa, and D.~Manzano, \textit{Configuration complexities of hydrogenic atoms}, Eur.~Phys.~J.~D {\bf 55}, 539 (2009).
	\bibitem{3jsymbols} L.~C.~Biedenharn and J.~D.~Louck, \textit{Angular Momentum in Quantum Physics} (Addison-Wesley, Reading) (1981). 
	
	
	
	
	\end{thebibliography}
\end{document}